\documentclass[journal]{IEEEtran}
\usepackage{graphics}
\usepackage{amsmath}
\usepackage{import}
\usepackage{times}

\usepackage{soul}
\usepackage{url}
\usepackage{algorithm}
\usepackage{algpseudocode}
\usepackage[hidelinks]{hyperref}
\usepackage[utf8]{inputenc}
\usepackage[small]{caption}
\usepackage{graphicx}
\usepackage{makecell}
\usepackage{amsmath}
\usepackage{booktabs}
\usepackage{import}
\usepackage{multirow, multicol}
\usepackage{subcaption}
\usepackage{array}
\usepackage{comment}
\usepackage{color, colortbl}
\usepackage{amsfonts}
\usepackage[dvipsnames]{xcolor}
\definecolor{Gray}{gray}{0.9}
\definecolor{green}{rgb}{0.4, 1.0, 0.0}

\ifCLASSINFOpdf
\else
\fi
\begin{document}
%
\title{Symbolic Music Structure Analysis with Graph Representations and Changepoint Detection Methods}
%
%
%

\author{Carlos~Hernandez-Olivan,~Sonia~Rubio~Llamas,~and~Jose~R.~Beltran
\thanks{Corresponding author: Carlos Hernandez-Olivan, Department
of Electronic Engineering and Communications, University of Zaragoza, Zaragoza 50018 Spain.\\ e-mail: carloshero@unizar.es.}
\thanks{}
\thanks{Preprint.}}

%
%

\markboth{Preprint}%
{Shell \MakeLowercase{\textit{et al.}}: Bare Demo of IEEEtran.cls for IEEE Journals}
%



\maketitle

\begin{abstract}
Music Structure Analysis is an open research task in Music Information Retrieval (MIR). In the past, there have been several works that attempt to segment music into the audio and symbolic domains, however, the identification and segmentation of the music structure at different levels is still an open research problem in this area.
In this work we propose three methods, two of which are novel graph-based algorithms that aim to segment symbolic music by its form or structure: Norm, G-PELT and G-Window. We performed an ablation study with two public datasets that have different forms or structures in order to compare such methods varying their parameter values and comparing the performance against different music styles. We have found that encoding symbolic music with graph representations and computing the novelty of Adjacency Matrices obtained from graphs represent the structure of symbolic music pieces well without the need to extract features from it. We are able to detect the boundaries with an online unsupervised changepoint detection method with a $F_1$ of 0.5640 for a 1 bar tolerance in one of the public datasets that we used for testing our methods. We also provide the performance results of the algorithms at different levels of structure, high, medium and low, to show how the parameters of the proposed methods have to be adjusted depending on the level.
We added the best performing method with its parameters for each structure level to musicaiz, an open source python package, to facilitate the reproducibility and usability of this work.
We hope that this methods could be used to improve other MIR tasks such as music generation with structure, music classification or key changes detection.
\end{abstract}


%
\IEEEpeerreviewmaketitle

\section{Introduction} \label{sec:intro}
%
%
%
%
\IEEEPARstart{M}{usic} Structure (or \textit{form}) Analysis (MSA) is a field of the Music Information Retrieval (MIR) which consists on predicting the structure or \textit{form} of music pieces. Music structure is a music principle \cite{review} that is closely related to other music principles like the harmony. Music presents a hierarchical structure in which sections are ordered in a coherent way. Such sections contain different rhythmic patterns and harmonic progressions that express different ideas. However, in Western classical music such sections can be connected between themselves with the so called cadences or bridges. From the music cognition perspective, there are different ways of understanding the hierarchical structure in music: grouping structure, metrical structure, time-span reduction, and prolongational \cite{lerdahl1983overview}.

Attending to computational approaches of MSA, there are different subtasks that have been previously studied like \textit{melodic segmentation}, \textit{motif discovery} and \textit{structural segmentation} \cite{lopez2012automatic}. Structural segmentation can be also divided in two smaller tasks: \textit{boundary detection} which aims to segment a piece by its structure boundaries, and \textit{segment labeling} which aims to label or name the sections of a piece. In this paper we focus on the task of \textit{boundary detection}, as it is the first step in analyzing the form of a music piece.

The MSA has been studied from two different perspectives according to the nature of the input music: the symbolic domain and the audio domain. Whereas the symbolic domain refers to the music data that contain the basic information about the notes, instruments, etc (MIDI or MusicXML files), the audio domain refers to the digital musical signal and transforms that can be computed from it. This makes the MSA task harder in the audio domain since there is no high-level musical information in the digitized music signal. In both domains, the first step usually consists on predicting the boundaries or segmenting music, and the second step (if applies) labels the segments according to their similarities.

In the symbolic domain, segmentation has been studied for decades \cite{lopez2012automatic} specially regarding monophonic melodic segmentation \cite{rodriguez2014comparing}, \cite{lattner2015probabilistic}, \cite{bassan2022unsupervised}.

MSA present the following challenges:
\begin{itemize}
    \item Each music genre or style has a different form or structure, even in the same music genre there might be multiple forms.
    \item Each piece (even pieces of the same genre) has a different section lengths, number of themes and/or musical phrases.
    This means that we do not now \textit{a priori} the number of sections or themes in a piece. In addition, this problem leads to non-balanced data.
    \item The musical content of the sections with the same name in different pieces is very different, as the nomenclature is intrinsic to each piece. This makes it difficult to train a supervised approach that labels sections by their name.
    \item Boundaries represent small percentage of the data if we compare them to the number of notes in a piece, which makes this and the previous point difficult when it comes to train deep neural networks for section segmentation and/or labelling.
    \item There are different structure levels in music, e.g., in Western classical music form we can find the following levels: \textit{high} structure  (sections), \textit{mid} structure (themes or music phrases) and \textit{low} structure (motifs). However, not all the music can be divided in those levels, specially non-Western classical music.
    \item There are only a few datasets in MIR of symbolic music that contain structure annotations.
\end{itemize}

MSA not only covers the task of music analysis but it can also be used in music generation systems to reinforce such models to generate coherent music \cite{review}, \cite{arxiv.2205.08579}.

In this paper, we propose and compare three methods that predict the segment boundaries in symbolic music without any previous information about the number of boundaries in the pieces, nor the time signature or any other feature different from the note on, note off or pitch of every note in the piece. In addition, the methods are unsupervised and do not need a training algorithm. We validate our methods with two public multi-instrument datasets with different structures and music genres.

\subsection{Contributions}
\begin{figure}[!t]
    \centering
    \includegraphics[width=\columnwidth]{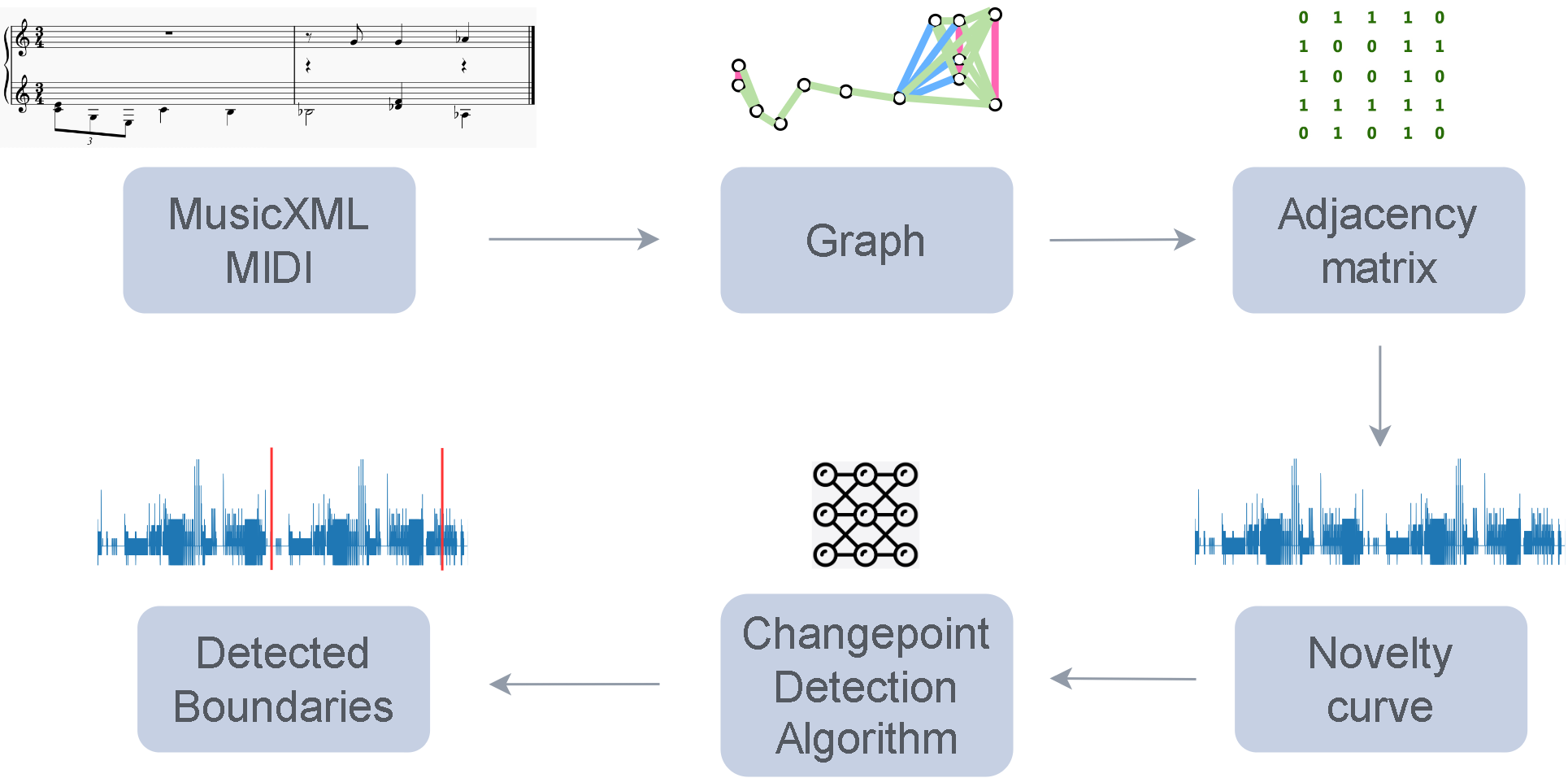}
    \caption{Our proposed method for symbolic music segmentation.}
    \label{fig:schema}
\end{figure}

The main contributions of this work are:
\begin{itemize}
    \item A novel fast online method that uses graphs to encode symbolic music which only needs temporal information for segmenting music by its structure (Figure \ref{fig:schema}).
    \item An ablation study about the segmentation of the three structure levels: low, mid and high.
    \item A comparison of the performance of the method with two different public datasets that contain different forms and instruments. Both datasets contain polyphonic multi-instrument music.
\end{itemize}

We make the code publicly available for reproducibility\footnote{\url{https://github.com/carlosholivan/symbolic-music-structure-analysis}, accessed February 2023}, and we describe some applications that can benefit from this work in Section \ref{sec:conclusions}.

\subsection{Paper Organization}
This paper is organized a follows: in Section \ref{sec:related} we discuss previous work done in MSA, in Section \ref{sec:boundaries} we describe the proposed methods, in Section \ref{sec:ablation} we compare them, in Section \ref{sec:eval} we evaluate our methods on two public datasets, in Section \ref{sec:discussion} we discuss the results and in Section \ref{sec:conclusions} we give conclusions and future work that could benefit from our work.

\section{Related Work} \label{sec:related}

\subsection{Audio Music Structure Analysis}
Music structure analysis have been studied in both the audio and symbolic domains \cite{msareview}. Music segmentation can be based on the following principles: homogeneity, repetition and novelty \cite{muller2015}. In the audio domain, one of the representations that is commonly used is the Self-Similarity Matrix (SSM) \cite{foote1999visualizing}. SSMs have been used as inputs of deep learning models in previous studies of boundary detection 
\cite{boundaries2021hernandezolivan}. Other works use variations of such matrices for unsupervised audio segmentation \cite{serra2012unsupervised}.

In Eq. \ref{eq:ssm} we show the expression of the SSM for a time series $Y = [y_1, ..., y_n] \in R^n$.

\begin{equation}
    \text{SSM}_{i, j} = distance(y_i, y_j)
    \label{eq:ssm}
\end{equation}

Also in the audio domain $l_1$-graph representations of audio features have been proposed for MSA \cite{PanagakisKA11}. $l_1$-graphs are a type of graphs that in which the edges between nodes are weighted based on the absolute difference between the values of their corresponding features. In this approach, audio features such as the \textit{Auditory Temporal Modulations} (ATMs),
the \textit{Mel-frequency Cepstral Coefficients} (MFCCs), and the
\textit{Chroma} features are used as inputs.

\subsection{Symbolic Music Structure Analysis}
In the symbolic domain, there have been proposed different techniques to address the melody segmentation and structure segmentation tasks.

Boundary detection in symbolic music has been studied for decades \cite{lopez2012automatic} mostly for monophonic melodic segmentation \cite{rodriguez2014comparing}, \cite{lattner2015probabilistic}, \cite{bassan2022unsupervised}. These methods can be applied to annotate large corpus of data such as \texttt{SIATEC} and \texttt{COSIATEC} algorithms \cite{meredith2002algorithms}.
The vast majority of these feature-based methods use rule-based systems or need to extract features such as the \textit{Inter Onset Intervals} (IOIs, Eq. \ref{eq:iois}) from the symbolic data to find a structural description of the music \cite{lerdahl1996generative}. Examples of that are the Local Boundary Detection Model (LBDM) \cite{Cambouropoulos01} that uses the IOIs to segment music phrases or the Grouper Program \cite{temperley2004cognition} which uses the \textit{Offset to Offset Intervals} (OOI) in addition to the IOIs. Extensions of these methods use variations of this features (e.g. local variations of IOIs) to perform the same task \cite{cenkerova2018crossing}. Similarly, the Pattern Boundary Detection Model (PAT) \cite{cambouropoulos2004influence} uses symbolic features to perform phrase segmentation. Adaptive melodic segmentation in MIDI files have also been proposed \cite{wilder2008adaptive}. 

Rule mining techniques \cite{Kranenburg20} and pseudo-supervised methods \cite{LattnerCG15} are other techniques to perform the melody segmentation task. For this purpose, there are annotated corpus \cite{LopezV13} like TAVERN \cite{tavern}. The dataset contains 27 sets of theme and variations for piano by Mozart and Beethoven.

Focusing on higher level structures, there have been proposed computational analysis techniques of different musical forms. From J. S. Bach fugues \cite{10.2307/43829264} to the Sonata form structure that has been studied from W. A. Mozart string quartets \cite{allegraud2019learning}. The techniques used in these works are mostly feature-based with harmonic and rhythmic features such as the rhythm break or the triple hammer blow.

\subsection{Symbolic Music as Graphs}
A graph $G(V, E)$ can be defined as a collection of points that have different types of connections. The points are called nodes $V$ and the connections between the nodes are the edges $E$. Graph representations are becoming popular in applications such as proteins discovery \cite{jumper2021highly} or to simulate complex physics \cite{physics}. In symbolic music, graph representations have been proposed for diverse purposes that concern MSA or music classification \cite{978-3-642-02124-4_5}.

Simonetta et al. \cite{SimonettaCOR18} used a graph-based representation to find similarities in symbolic music pieces. Jeong et al. \cite{jeong2019graph} used a graph enconding to learn note representations from music scores. Graphs have also demonstrated to be valid representations for Perfect Authentic Cadences (PAC) identification, which is closely related to structure segmentation. Karystianos and Widmer \cite{abs-2208-14819} proposed a stochastic Graph Convolutional Network (SGSMOTE) that adresses this problem in Bach Fugues, and Haydn and Mozart String Quartets. To construct the graph, notes and rests are nodes of the graph, and three types of undirected edges between nodes are defined as for consecutive notes or rests in time, for notes with the same onset and for notes whose onset happens when a longer note is already being played.

\section{Proposed methods}
\label{sec:boundaries}
Music Structure Analysis have been studied in the audio domain with the self-similarity matrices and derivations of them. In the symbolic domain, although boundary detection might seem simpler because we have basic information about notes such as pitch or note onset and offset, to our knowledge, there is not much work that has attempted to segment music structure at different levels. In this section, we propose three algorithms for music structure detection in symbolic music: Norm, G-PELT and G-Window. 

We implemented the Norm algorithm based on previous work. The method extracts the IOIs and pitch direction of the music to obtain the boundaries of the structure segments. G-PELT and G-Window are novel graph-based methods that segment different levels in the structure of symbolic music depending on the parameters values (see Table \ref{tab:optimal}) The algorithms do not need the symbolic music to be quantized, not the time signature nor the beats per minute information, which means that we can use both methods in non curated MIDI files or in musicXML files.

\subsection{Method 1: Norm}
This method is based on the normalization of the IOIs and pitch direction that gives us boundaries candidates that are used to construct a self-similarity-matrix. Since in symbolic music we do have more information about the notes than in audio signals, we construct the SSM by grouping the notes in segments and measuring the distances between the segments. In Algorithm \ref{alg:norm} we provide the pseudo-code of the Norm algorithm. The algorithm consists of the following steps:

The first step to identify the boundaries in a symbolic note sequence is to sort the notes by their \textit{onset} or \texttt{Note ON} (in MIDI notation) and compute the IOIs vector $\mathbf{x} \in R^{N-1}$ where $N$ is the number of notes in the music piece (Eq. \ref{eq:iois}):
\begin{equation} \label{eq:iois}
    x_i = \textit{onset}_{i+1} - \textit{onset}_{i} \quad i = 1 \ldots N
\end{equation}

We define the local direction or pitch contour vector $\mathbf{l} \in R^{N-1}$ similar to previous works in which the contour intervals were stored in an array of contour intervals called COM-matrix \cite{10.2307/745814}. In this work, we compute $\mathbf{l}$ as follows: if a note is followed by a note with a higher pitch (ascendent), $l_i=1$ and $l_i=-1$ if it is lower (descendent). If the pitch of the following note is equal to the current pitch then $l_i=0$. In Eq. \ref{eq:l} we show the expression of the local direction vector:

\begin{equation}
    l_i = \left\{ \begin{array}{lcc}
             1 &   \forall  & p_{i+1} > p_i 
             \\ -1 &   \forall  & p_{i+1} < p_i 
             \\ 0 &   \forall  & p_{i+1} = p_i 
             \end{array}
             \quad i = 1 \ldots N
   \right.
    \label{eq:l}
\end{equation}
where $p_i$ refers to the pitch of the $ith$ note.

After computing $\mathbf{x}$ and $\mathbf{l}$ we sum them to construct the vector $\mathbf{\tilde{x}}$. We normalize this vector to obtain the $\mathbf{\hat{x}}$ vector by applying the z-score normalization (Eq. \ref{eq:norm}):
\begin{equation} \label{eq:norm}
    \hat{x}_i = \frac{x_i - \mu }{\sigma} \quad i = 1 \ldots N
\end{equation}
where $\mu$ is the mean and $\sigma$ the standard deviation of the whole IOIs vector.

With $\mathbf{\tilde{x}}$, we now apply a peak picking strategy. By doing this, we extract the segment boundaries candidates $\mathbf{b} = [b_1, ..., b_S] \in R^C$ with $C$ the number of boundary candidates. We select a window size $w_1$ and a threshold $\tau_1$ (standard deviation above the mean). The window size $w_1$ is defined as a function of the number of notes $N$ so $w=\frac{\alpha}{\hat{n}}$ where $\hat{n}=15$ and $\alpha$ is a constant parameter. We fixed $\hat{n}=15$ to optimize the algorithm performance.

After that, we can compute the SSM, $\mathbf{S} \in R^C \times R^C$, by grouping the notes in the obtained boundary candidates. The distance to construct the $\mathbf{S}$
is the euclidean distance. Since not all the segments have the same length, we add zeros to the segments of lesser length to calculate the distance between all of them.

Once $\mathbf{S}$ is computed, we can get its novelty curve $\mathbf{c}$ (Eq. \ref{eq:nov}) and perform again the peak picking strategy with a sliding window $w_2$ and a threshold $\tau_2$. This will give us the predicted boundaries $\mathbf{\hat{b}}$. 
\begin{equation}
    c_i = norm(s_{i+1}, s_i)
    \label{eq:nov}
\end{equation}
where \textit{norm} refers to the euclidean norm.

\begin{figure}[!h]
\subfloat[Normalized IOIs and local direction vector with the boundary candidates detected after the peak picking (in red).\label{ssm:a}]{%
    \centering
    \includegraphics[width=\columnwidth]{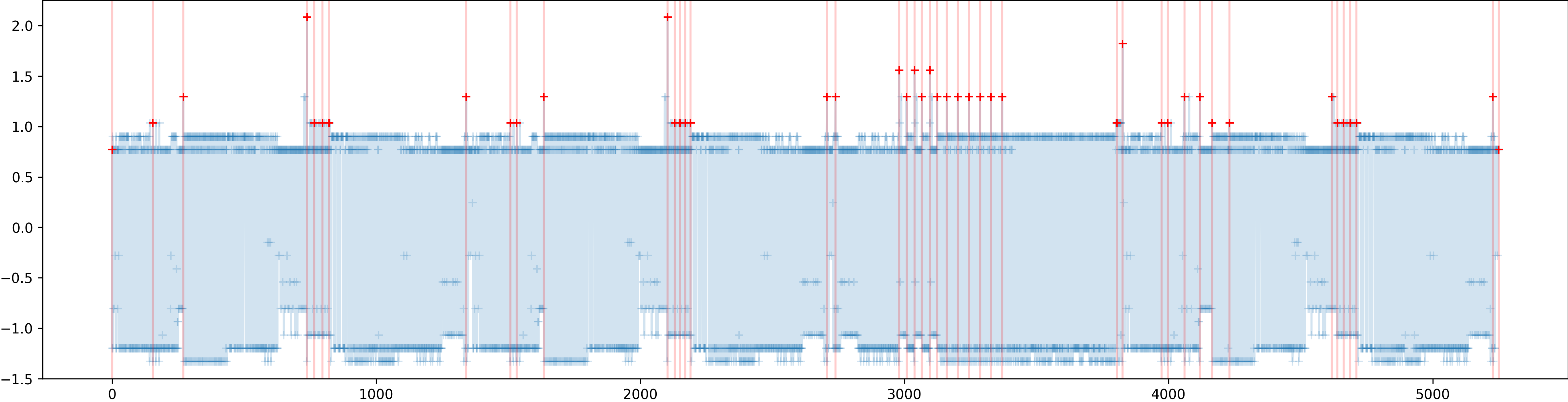}
} \\
\subfloat[SSM and novelty curve with predicted boundaries (in red)\label{ssm:b}]{%
    \centering
    \includegraphics[width=\columnwidth]{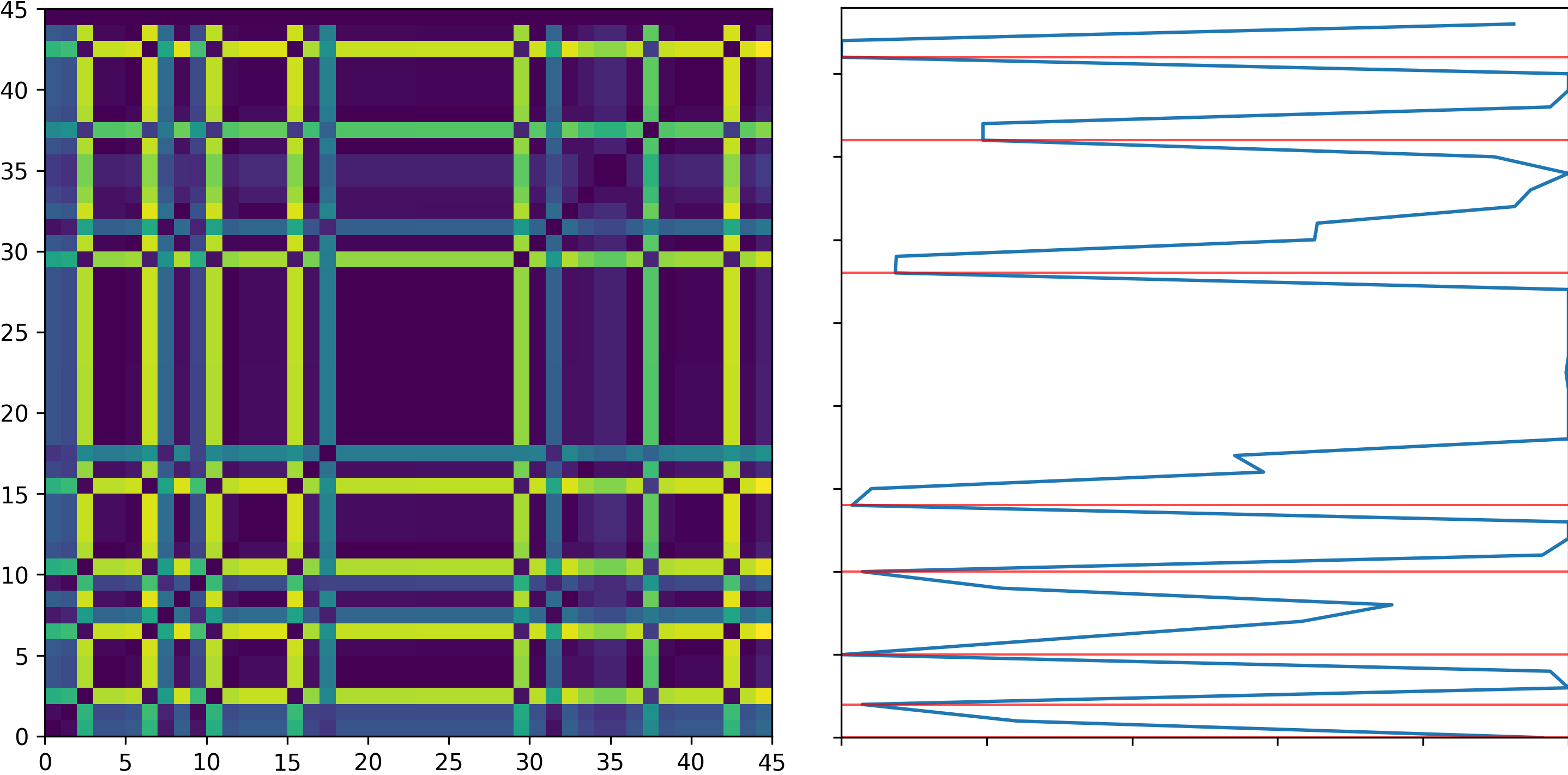}
}

\caption{Boundary candidates (a) and SSM with its novelty curve (b) in the sample nº1 of Beethoven Piano Sonatas Dataset.}
\label{fig:msa_norm}
\end{figure}

In Fig. \ref{ssm:a} we show the identified peaks (boundary candidates) in the normalized 
$\mathbf{\hat{b}}$ vector, and in Fig. \ref{ssm:b} the SSM with the novelty curve and the predicted boundaries for the Schubert\_D911-01.mid file of the Schubert Winterreise Dataset (SWD) .

The procedure of the Norm method is summed up below, and its pseudo-code can be found in Algorithm \ref{alg:norm}.

\begin{itemize}
    \item Calculate the IOI's $\mathbf{x}$ and the local direction $\mathbf{l}$ vectors of the file and sum them: $\mathbf{\tilde{x}}$.
    \item Apply z-score normalization to $\mathbf{\tilde{x}}$: $\mathbf{\hat{x}}$.
    \item Calculate the boundary candidates, $\mathbf{b}$, by peak picking with window $w_{1}$ and threshold $\tau_{1}$.
    \item Construct a Self-Similarity Matrix, $\mathbf{S}$, with the segments grouped by the boundary candidates obtained in the previous step.
    \item Get the novelty curve $\mathbf{c}$ from $\mathbf{S}$.
    \item Calculate the boundaries  $\mathbf{\hat{b}}$ by peak picking with window $w_{2}$ and threshold $\tau_{2}$.
\end{itemize}

\begin{algorithm}
    \caption{Norm}\label{alg:norm}
    \begin{algorithmic}
    \Require $w_{1}, \tau_{1}, w_{2}, \tau_{2}$ \Comment{params}
    \State $n \gets read\_file$
    \State \Comment{calculate candidates $b$}
    \State $x \rightarrow ioi(n)$
    \State $l \rightarrow local\_direction(n)$
    $\hat{x} \gets sum(x, l )$
    \State $z = z\_normalization(\hat{x})$
    \State $b \rightarrow peak\_picking(z, w_{1}, \tau_{1})$
    
    \State \Comment{group notes by candidates $b$}
    \State $g \gets$ empty list of lists
    \For{$i$ in $b$} 
    \State $g \gets append(n_i, n_{i+1})$
    \EndFor
    
    \State \Comment{construct $S$}
    \State $S \gets array_{len(b), len(b)}$
    \For{$i$ in $g$}
    \State $v_i \gets sum(ioi(g_i) \lvert\lvert local\_direction(g_i))$
    \For{$j$ in $g$}
    \State $v_j \gets sum(ioi(g_j) \lvert\lvert local\_direction(g_j))$
    \State $S_{i,j} = euclidean\_dist(v_i, v_j)$
    \EndFor \EndFor
    \State $c \gets novelty(S)$
    
    \State \Comment{get $b\prime$ and the respective note positions}
    \State $b^\prime \rightarrow peak\_picking(n, w_{2}, \tau_{2})$
    \For{$k$ in $b^\prime$}
    \State $b^\prime\_notes \gets append(b_k)$
    \EndFor
    \end{algorithmic}
\end{algorithm}

\subsection{Graph Representation} \label{sec:graph}
Our graph representation is based on previous works \cite{abs-2208-14819}, \cite{jeong2019graph}, \cite{szeto2006graph}. Since our representation is based on MIDI files, which might do not contain time signature and tempo information, neither bars nor beats will be taken into account in our representation. We only extract the time signatures from the datasets metadata for measuring the performance of our methods in Section \ref{sec:ablation}. Similar to \cite{abs-2208-14819}, we create three types of undirected connections
only between notes (nodes), however we do not encode rests since MIDI files do not provide them directly. In Eq. \ref{eq:edges} we can see the expression of the edges of our graph representation: $E^{\textit{on}}$ represent the edges between notes that occur on the same onset; $E^{\textit{cons}}$ are the edges between consecutive notes in time,
and edges $E^{h}$ represent overlapping notes in time (notes that start when there is already a note being played). The edges of the graph $E$ are defined as $E \subseteq E^{\textit{on}} \cup E^{\textit{cons}} \cup E^{h}$:

\begin{equation}
\begin{split}
    &e^{\textit{on}}_{i,j} =  \{ \textit{on}(v_i) = \textit{on}(v_j) \}
    \\
    &e^{\textit{cons}}_{i,j} = \{\textit{off}(v_i) = \textit{on}(v_j) \}
    \\
    &e^{h}_{i,j} =  \{ [\textit{off}(v_i) > \textit{on}(v_j)] \wedge [\textit{on}(v_i) < \textit{on}(v_j) ] \}
\end{split}
\label{eq:edges}
\end{equation}
where \textit{on} and \textit{off} refer to the notes \textit{onsets} and \textit{offsets}, respectively.

\begin{figure}[!h]
\includegraphics[width=\columnwidth]{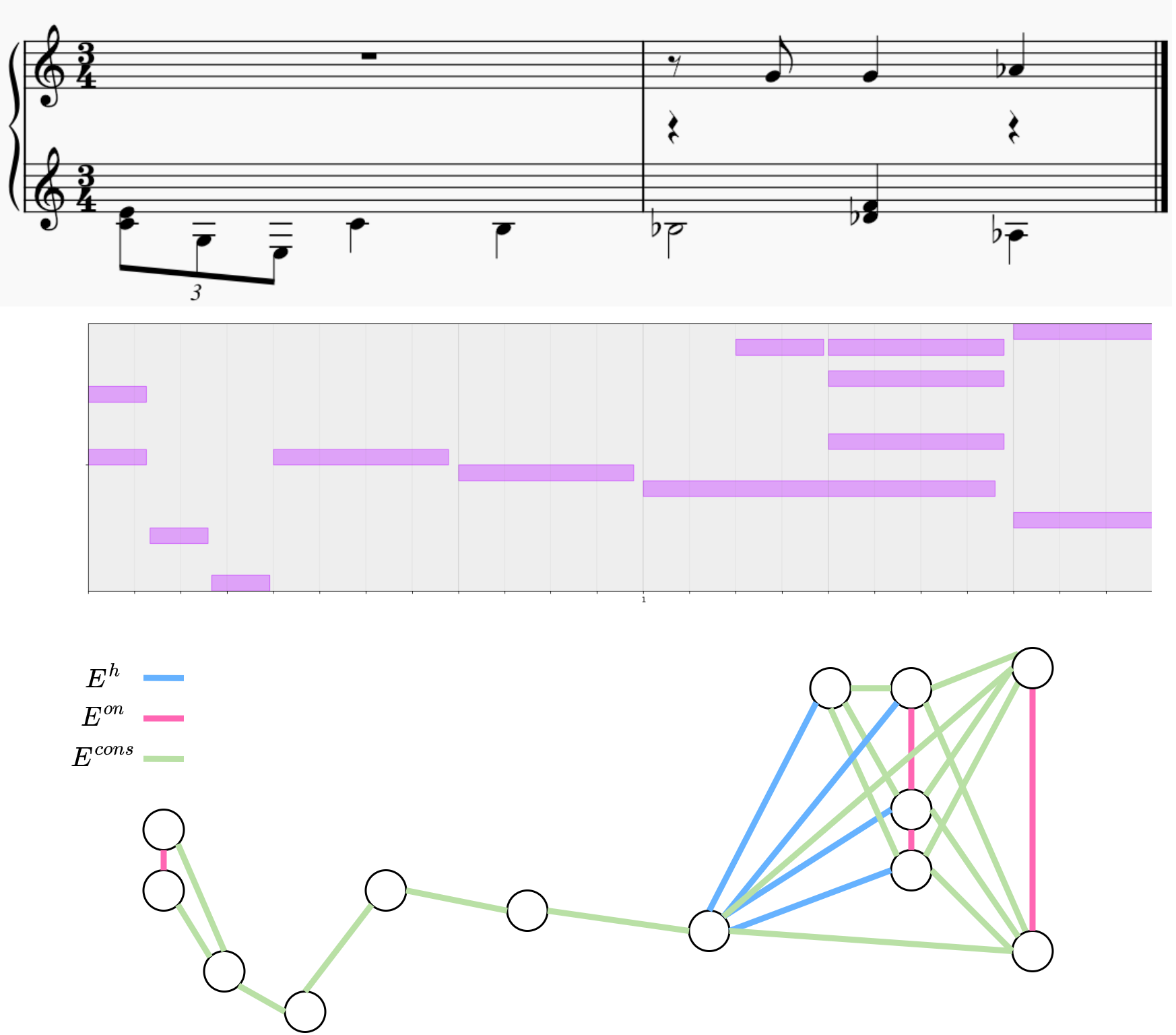}
\caption{From top to down: score, pianoroll and our proposed graph representation of 2 bars of symbolic music.}
\label{fig:repr}
\end{figure}

Nodes are not connected between themselves. In Fig. \ref{fig:repr} we show an example of 2 bars in a score, pianoroll and our graph representation.

\subsubsection{PELT Algorithm}
In this method, we encode symbolic music as a graph as we described previously to then obtain the adjacency matrix $\mathbb{A}$ of the graph. Note that constructing $\mathbb{A}$ has a quadratic complexity with the number of notes $n$ in the file: $\mathcal{O}(n^2)$. We then compute the novelty curve $\mathbb{c}$ of $\mathbb{A}$ with the expression in Eq. \ref{eq:nov}. After that, we run the PELT algorithm with \texttt{ruptures} package\footnote{\url{https://github.com/deepcharles/ruptures}, accessed February 2023.} \cite{TruongOV20}.

PELT \cite{pelt} is a changepoint detection algorithm in time series. It detects the change points of a signal by optimizing a cost function $C$ as defined in Eq. \ref{eq:cost}. 

To calculate the optimal changepoint location, the algorithm starts initializing the cost of the entire series $C_i$ to infinity, then calculates the cost of each possible changepoint location per step, selects the point $i$ that minimizes the cost and then updates the cost function with the new location $i^\prime$. In Eq. \ref{eq:cost} we show the expressions that correspond to this process. After that, the windows are reduced in size by a factor $\delta$ and the process in Eq. \ref{eq:cost} is repeated.

\begin{equation}
\begin{gathered}
   C_i = min(C_j) + \lambda(i, j) + p
    \\
    i^\prime = argmin(C_i)
    \\
    C_i = min(C_i, C_{i^\prime})
\end{gathered}
\label{eq:cost}
\end{equation}
where $i, j$ is the segment to be processed, and $p$ is the penalty, which controls the number of the detected changepoints. 

The window size $w$ controls the size of the analyzed segments per iteration, and the jump value $j$ is related to the number of changepoints skipped in each iteration. We also defined in this case the window size $w$ as a function of the number of notes $N$ so $w=\frac{\alpha}{N}$ where $N=15$ and $\alpha$ is a constant parameter. We define the jump value $j=\beta \cdot w$  as a function of the window size $w$ and a constant $\beta$. We find the optimal values of parameters $\alpha$ and $\beta$ in Section \ref{sec:ablation}.

The complexity of the method depends on the cost function used. In our case, we use the Kernelized mean change cost function RBF of Eq. \ref{eq:cost_f} which has a quadratic complexity corresponding to the sum of the computation of the kernel density estimate (KDE),  $\mathcal{O}(n^2)$, and the computation of the cost at each candidate is $\mathcal{O}(n)$. 
This leads to an overall complexity of the PELT method of $\mathcal{O}(n^2 + n log(n))$, since the number of iterations is determined by the logarithm of the ratio of the window size and the minimum size $\mathcal{O}(log(n))$.

\begin{equation}
    C(\theta) = \sum_{t=1}^{T} \left\lVert y_t - \sum_{j=1}^{k} w_j \phi\left(\frac{\left\lVert x_t - \mu_j\right\rVert^2}{\sigma^2}\right)\right\rVert^2
    \label{eq:cost_f}
\end{equation}
where $C(\theta)$ is the cost function,
$\theta = {w_j, \mu_j, \sigma^2}$ is a set of parameters,
$T$ is the number of points in the signal,
$y_t$ is the observed value at time $t$,
$x_t$ is the feature value at time $t$,
$w_j$ is the weight for the $j^{th}$ RBF component,
$\mu_j$ is the mean of the $j^{th}$ RBF component,
$\sigma^2$ is the variance of the RBF components, and
$\phi(u) = \exp(-u)$ is the RBF function
$\left\lVert \cdot \right\rVert$ is the Euclidean norm.

\begin{figure*}[!t]
\centering
    \includegraphics[width=\textwidth]{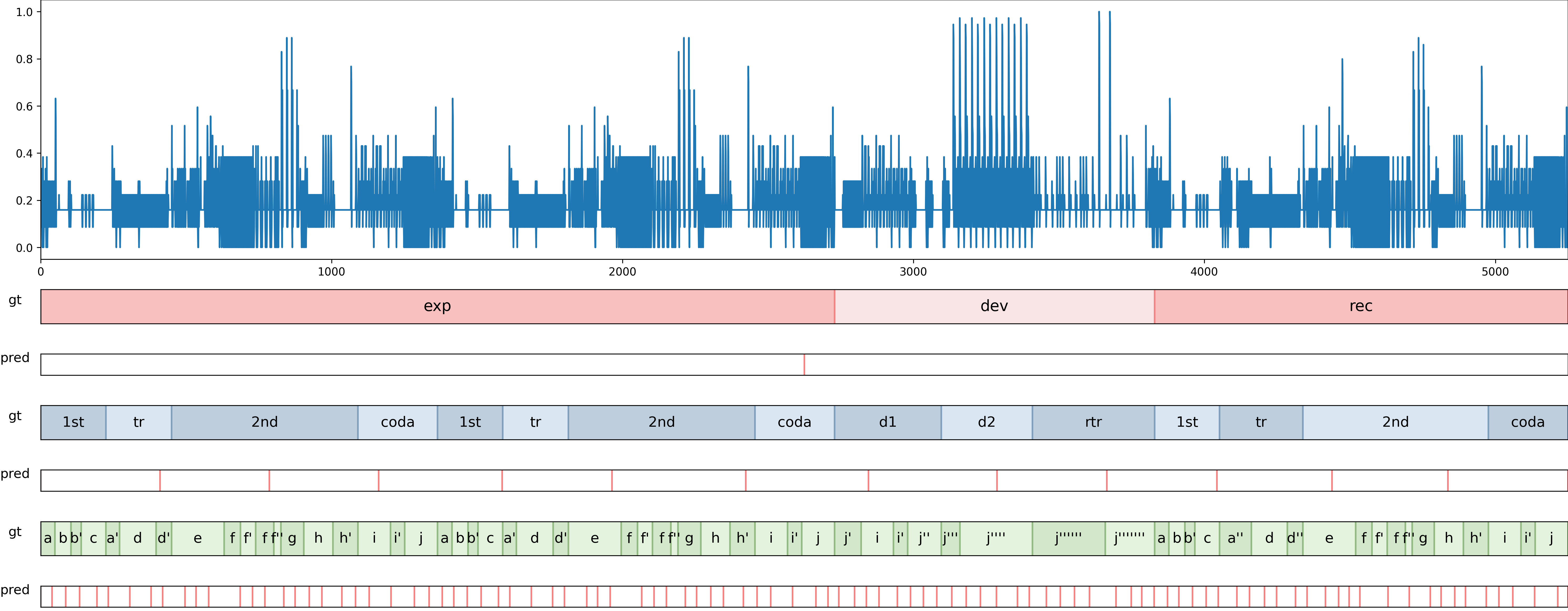}
    \caption{Adjacency matrix novelty curve of the graph built of the Beethoven piano sonata n11, 1st movement. In the figure, gt and pred refers to the ground truth annotations and predicted boundaries with out G-PELT method, respectively. The x axis refers to the notes in the file and the y axis is the value of the novelty.}
\label{fig:msa-graph}
\end{figure*}

In Fig. \ref{fig:msa-graph} we can see the predicted boundaries in a sample of the Shuber Winterreise Dataset (SWD). We can observe how the algorithm predicts different boundaries at different levels of structure.

In Algorithm \ref{alg:g-pelt} we provide the pseudo-code of the PELT algorithm re-adapted for our graph representation and renamed to G-PELT.

\begin{algorithm}
    \caption{G-PELT}\label{alg:g-pelt}
    \begin{algorithmic}
    \Require w, j, p \Comment{params}
    \State $n \gets read\_file$
    \State $G \gets midi\_to\_graph(n)$
    \State $A \gets adjacency\_matrix(G)$
    \State $c \gets novelty(A)$
    \State $b^\prime \gets$ PELT$(n, w, j, p)$
    \end{algorithmic}
\end{algorithm}

\subsubsection{Window Algorithm}
In this method, we also encode symbolic music as a graph and compute the novelty $\mathbf{c}$ from the adjacency matrix $\mathbf{A}$ of the graph as we do in the G-PELT algorithm.
After that, we applied a sliding window algorithm. The algorithm uses two windows $y_{i, j}$ and $y_{j, k}$, that are compared by computing a discrepancy measure with the cost function $C$ of each window as we show in Eq. \ref{eq:window}.

\begin{equation}
    d(y_{i,j}, y_{j,k}) = C(y_{i,j}) - C(y_{i,k}) - C(y_{j,k})
    \label{eq:window}
\end{equation}

The window size $W$ defines the segment splits for each signal point $i$ as: $y_{i-w/2, i}, y_{i, i+w/2}$.

The computational complexity of the algorithm is
$\mathcal{O}(nw)$. We need to add the complexity of the cost function $C$. We use the same function defined for G-PELT algorithm (see Eq. \ref{eq:cost_f}), which computational complexity in this case is $\mathcal{O}(nw)$ since the complexity of computing the KDE cost function for a single window is $\mathcal{O}(w^2)$ and since the window is moved along the data, the total time required to compute the cost function for the entire data set is $\mathcal{O}(nw)$. This leads to an overall complexity of the sliding window algorithm to $\mathcal{O}(2nw)$.

In Algorithm \ref{alg:g-clasp} we provide the pseudo-code of the G-Window algorithm and in Table \ref{tab:algs} we show the proposed algorithms for the symbolic music segmentation task with their parameters and complexity.

\begin{algorithm}
    \caption{G-Window}\label{alg:g-clasp}
    \begin{algorithmic}
    \Require w, n\_p \Comment{params}
    \State $n \gets read\_file$
    \State $G \gets midi\_to\_graph(n)$
    \State $A \gets adjacency\_matrix(G)$
    \State $c \gets novelty(A)$
    \State $b^\prime \gets$ Window$(n, w, n\_p)$
    \end{algorithmic}
\end{algorithm}

\begin{table}[!h]
\caption{Parameters and complexity of the algorithms for symbolic music structure segmentation. Note that the complexity of constructing the adjacency matrix is separated from the complexity of PELT and Window algorithms.}
    \centering
    \begin{tabular}{
    p{1.5cm}
    >{\centering\arraybackslash}p{2cm}
    >{\centering\arraybackslash}p{3.5cm}
    }
    \toprule
         Algorithm & Params. & Complexity\\
         \hline
        Norm & $\alpha_{1}, \tau_{1}, w_{2}, \tau_{2}$ & $\mathcal{O}(n) + \mathcal{O}(b^2) $ \\
        G-PELT & $\alpha, \beta, p$ & $\mathcal{O}(n^2)$ + $\mathcal{O}(n^2 + n log(n))$ \\
        G-Window & $\alpha, \beta, p$ & $\mathcal{O}(n^2)$ + $\mathcal{O}(2nw)$ \\
        \bottomrule
    \end{tabular}
    \label{tab:algs}
\end{table}

\section{Ablation Study} \label{sec:ablation}
In this section we perform an ablation study to compare the methods described in Section \ref{sec:boundaries} with different parameter values. The goal of this comparison is to find the optimal parameter values for each method and dataset, and to determine the method that outperforms the others.

As we mentioned in Section \ref{sec:boundaries}, our representation is based on MIDI files where no information about bars nor beats is provided. This makes the algorithms more robust since MIDI files may not be quantized, especially files containing expressive performances. Thus, it is easier to convert a MusicXML file to MIDI and use our method for score segmentation rather than converting a MIDI file to MusicXML due to the ornamentations and the lack of symbolic information that a MIDI file provides in comparison to a score.

Being that said, we convert MIDI files into graphs. The implementation has been added to \texttt{musicaiz} package \cite{musicaiz} which uses \texttt{NetworkX} package\footnote{\url{https://github.com/networkx/networkx}, accessed November 2022.} \cite{networkx} to do the conversion. We test the algorithms performance with \texttt{mir\_eval} package \cite{raffel2014mir_eval}.

For testing, we select 2 tolerances: 1 beat and 1 bar tolerances. The 1 beat tolerance has been used previously in the cadences identification task \cite{abs-2208-14819}. We added the 1 bar tolerance to give further insights about the performance of the algorithms \footnote{Note that our tolerance values are an analogy to the boundaries detection task in the audio domain, where there are 2 tolerance values 0.5 and 3 seconds. 1 beat tolerance means that in a 3/8 bar, the beat will be the crotchet }.

\subsection{Datasets}
We use the Schubert Winterreise Dataset (SWD) which contains 24 files (MIDI, MusicXML, PDF and audios) annotated with harmony and form analysis (our task).
Since we aim to test our methods for different music forms, we also test the algorithms with the Beethoven Sonatas Dataset (BPS) which contains the annotations of the 32 first movements of Beethoven piano sonatas. However, this dataset provides the symbolic music files as \texttt{csv}, so we converted them to MIDI format with \texttt{musicaiz} package to be able to read, process and measure them against the structure annotations provided by the dataset. In Table \ref{tab:data} we show the metadata of both datasets\footnote{We exclude file 11 since it contained tempo changes}. Whereas SWD has only annotations of the structure in the middle level, BPS dataset has annotations for the three levels that we called: low, mid and high.

\begin{table}[!t]
    \normalsize
    \centering
    \caption{Datasets analytics.}
    \begin{tabular}{
    p{1.2cm}|  
    >{\centering\arraybackslash}p{0.4cm}|
    >{\centering\arraybackslash}p{1.2cm}|
    >{\centering\arraybackslash}p{1.5cm}|
    >{\centering\arraybackslash}p{2.4cm}
    }
    \hline
    Dataset & files & TSig: files & Total Bound. & Boundaries per file\\
    \cmidrule{1-5}
    \multirow{6}{*}{SWD} & \multirow{6}{*}{23} & 2/4: 7 & \multirow{6}{*}{mid: 192} & \multirow{6}{*}{mid: $8.41_{\pm 2.79}$}\\
     & & 3/4: 7 & & \\
     & & 4/4: 4 & & \\
     & & 3/8: 1 & & \\
     & & 6/8: 3 & & \\
     & & 12/8: 1 &  & \\
    \cmidrule{1-5}
    \multirow{6}{*}{BPS} & \multirow{6}{*}{31} & 2/4: 8 & \multirow{2}{*}{low: 1.439} & \multirow{2}{*}{low: $46.42_{\pm 21.55}$}\\
     & & 3/4: 6 & & \\
     & & 4/4: 13 & \multirow{2}{*}{mid: 438} & \multirow{2}{*}{mid: $14.13_{\pm 3.46}$}\\
     & & 3/8: 1 & & \\
     & & 6/8: 2 & \multirow{2}{*}{high: 115} & \multirow{2}{*}{high: $3.8_{\pm 1.05}$}\\
     & & 12/8: 1 &  & \\
    \bottomrule
    \end{tabular}
    \label{tab:data}
\end{table}

To compare the performance of our methods, we set the lower bound (renamed to baselines from now on) for each dataset following the procedure introduced in previous works \cite{grill2015music}. This limit is the worst case scenario since boundaries are placed randomly in each file. For the SWD dataset we set 5 synthetic equidistant boundaries per file and in the BPS dataset we set 4, 14 and 46 boundaries per file which correspond to the mean of the boundaries per file and level (see Table \ref{tab:data}).

\subsection{Schubert Winterreise Dataset (SWD)}
In this subsection, we test the performance of the algorithms in the SWD. To be able to process the information and convert the symbolic information to graphs, we use \texttt{musicaiz} package \cite{musicaiz} which works only with MIDI files, thus, we use the raw MIDI files of SWD.

\subsubsection{Norm}
\begin{figure*}[!ht]
\subfloat[Norm 1 beat with $\alpha_1$ variation and $\tau_1$=1, $w_2$=2, $\tau_2$=0.5. \label{fig:norm_swd_a}]{%
    \centering
    \includegraphics[width=\columnwidth]{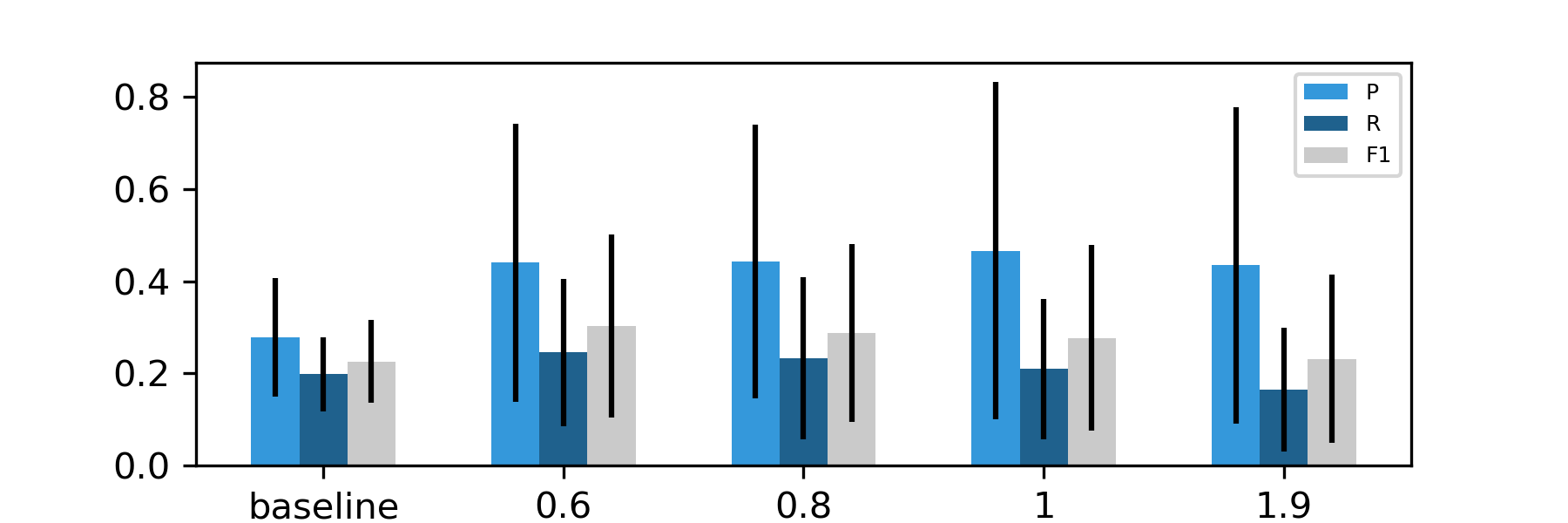}
}
\subfloat[Norm 1 bar with $\alpha_1$ variation and $\tau_1$=1, $w_2$=2, $\tau_2$=0.5. \label{fig:norm_swd_b}]{%
    \centering
    \includegraphics[width=\columnwidth]{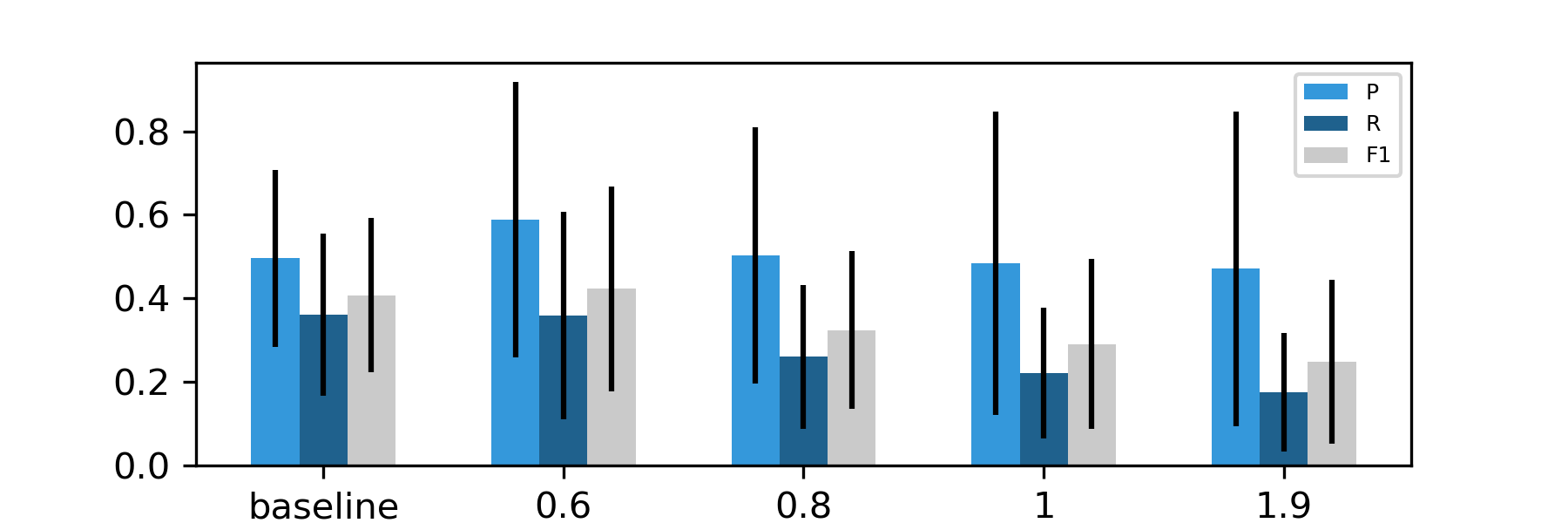}
} \\
\subfloat[Norm 1 beat with $\alpha_1$ variation and $\tau_1$=1, $w_2$=2, $\tau_2$=0.5. \label{fig:norm_swd_c}]{%
    \centering
    \includegraphics[width=\columnwidth]{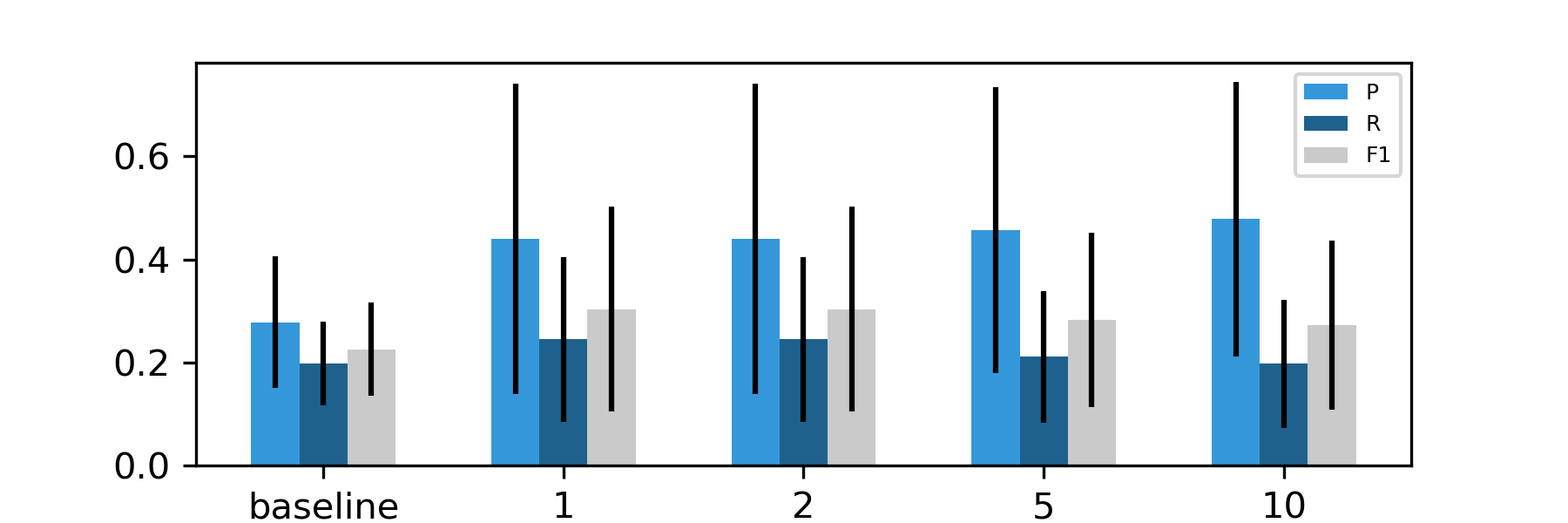}
}
\subfloat[Norm 1 bar with $w_2$ variation and $\tau_1$=1, $\alpha_1$=0.6, $\tau_2$=0.5.  \label{fig:norm_swd_d}]{%
    \centering
    \includegraphics[width=\columnwidth]{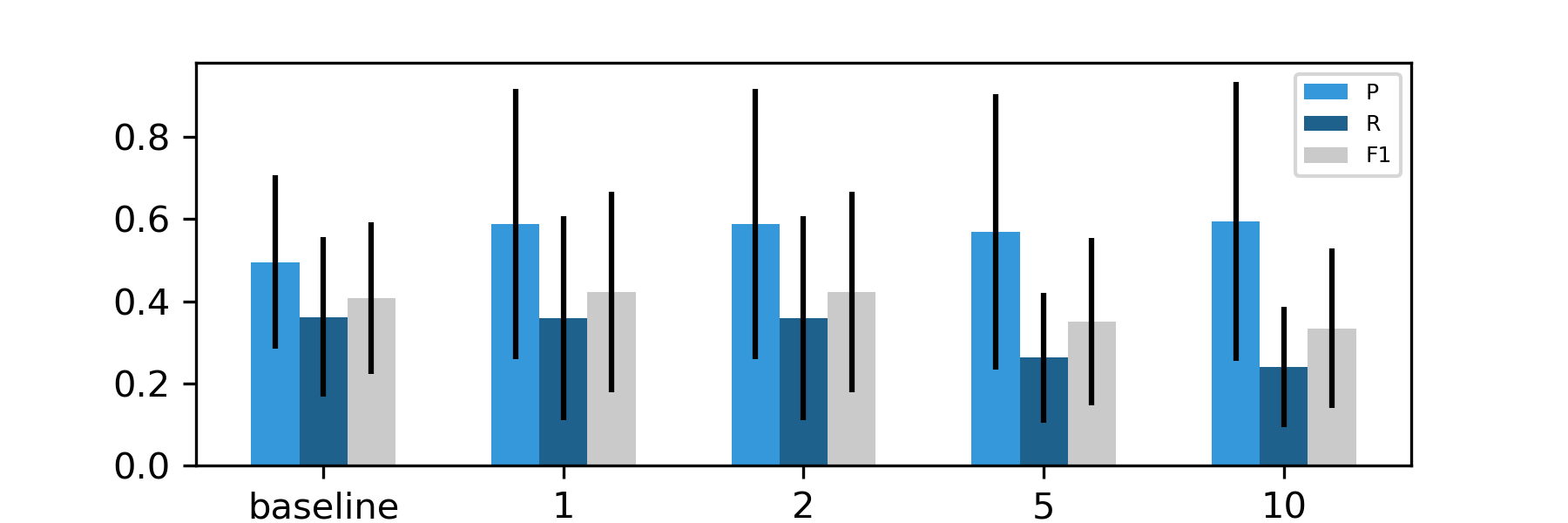}
} \\
\caption{Comparison of Norm method in the SWD dataset setting different $\alpha_1$ and $w_2$ values for 1 beat and 1 bar tolerances.}
\label{fig:norm_w}
\end{figure*}

We test the Norm algorithm varying its parameters $\alpha_1$, $\tau_1$, $w_2$, and $\tau_2$ in order to find their optimal values. In Fig. \ref{fig:norm_swd_a} we show the performance of the Norm method varying $\alpha_1$ and in Fig. \ref{fig:norm_swd_c} we vary $w_2$ with the optimal $\alpha_1$ value for 1 beat tolerance. In Figs. \ref{fig:norm_swd_b} and \ref{fig:norm_swd_d} we repeat the same procedure but testing the algorithm with 1 bar tolerance.

The results show that there is no a huge variation of the metrics when measuring the results with 1 beat tolerance, however, when testing the results with 1 bar tolerance the sensibility of the results against $\alpha$ and $w_2$ increases. The best performing parameters for both 1 beat and 1 bar tolerances are $\alpha$=0.6,  $w_2$=2, $\tau_1$=1 and $\tau_2$=0.5 with a $F_1$=0.3029 for 1 beat and $F_1$=0.4228 for 1 bar tolerances (see Table \ref{tab:msa}).

\begin{table}[!t]
    \centering
    \caption{MSA results for the algorithms with optimal parameters and measured with SWD.}
    \begin{tabular}{
    p{1.4cm}|  
    >{\centering\arraybackslash}p{1.5cm}
    >{\centering\arraybackslash}p{1.5cm}
    >{\centering\arraybackslash}p{1.5cm}
    }
    \cmidrule{1-4}
    & \multicolumn{3}{c|}{SWD}\\
    \hline
    & P & R & $F_1$\\
    \cmidrule{1-4}
        \rowcolor{Gray}
        & \multicolumn{3}{c|}{Tolerance: 1 beat}\\
        \cmidrule{1-4}
        baseline & $0.2782_{\pm 0.12}$ & $0.1980_{\pm 0.08}$ & $0.2258_{\pm 0.08}$\\
        Norm $\mathbf{b}$ & $0.2017_{\pm 0.14}$ & $0.4352_{\pm 0.31}$ & $0.2556_{\pm 0.17}$\\
        Norm $\mathbf{b}^\prime$ & $0.4398_{\pm 0.30}$ & $0.2450_{\pm 0.15}$ & $0.3029_{\pm 0.19}$\\
        G-PELT & $0.2665_{\pm 0.15}$ & $0.5265_{\pm 0.28}$ & $\mathbf{0.3455_{\pm 0.18}}$\\
        G-Window & $0.3066_{\pm 0.21}$ & $0.3372_{\pm 0.29}$ & $0.2869_{\pm 0.20}$\\
        \cmidrule{1-4}
        \rowcolor{Gray}
        & \multicolumn{3}{c|}{Tolerance: 1 bar}\\
        \cmidrule{1-4}
        baseline & $0.4956_{\pm 0.21}$ & $0.3619_{\pm 0.19}$ & $0.4078_{\pm 0.18}$\\
        Norm $\mathbf{b}$ & $0.3415_{\pm 0.18}$ & $0.7199_{\pm 0.31}$ & $0.4298_{\pm 0.18}$\\
        Norm $\mathbf{b}^\prime$ & $0.5884_{\pm 0.32}$ & $ 0.3588_{\pm 0.24}$ & $ 0.4228_{\pm 0.24}$\\
        G-PELT & $0.4366_{\pm 0.13}$ & $0.8473_{\pm 0.17}$ & $\mathbf{0.5640_{\pm 0.12}}$\\
        G-Window & $0.5371_{\pm 0.23}$ & $0.5527_{\pm 0.29}$ & $0.4863_{\pm 0.19}$\\
         \bottomrule
    \end{tabular}
    \label{tab:msa}
\end{table}

\subsubsection{G-PELT}
\begin{figure*}[!ht]
\subfloat[G-PELT 1\_beat with $\alpha$ variation and $\beta$=0.15.\label{1a}]{%
    \centering
    \includegraphics[width=\columnwidth]{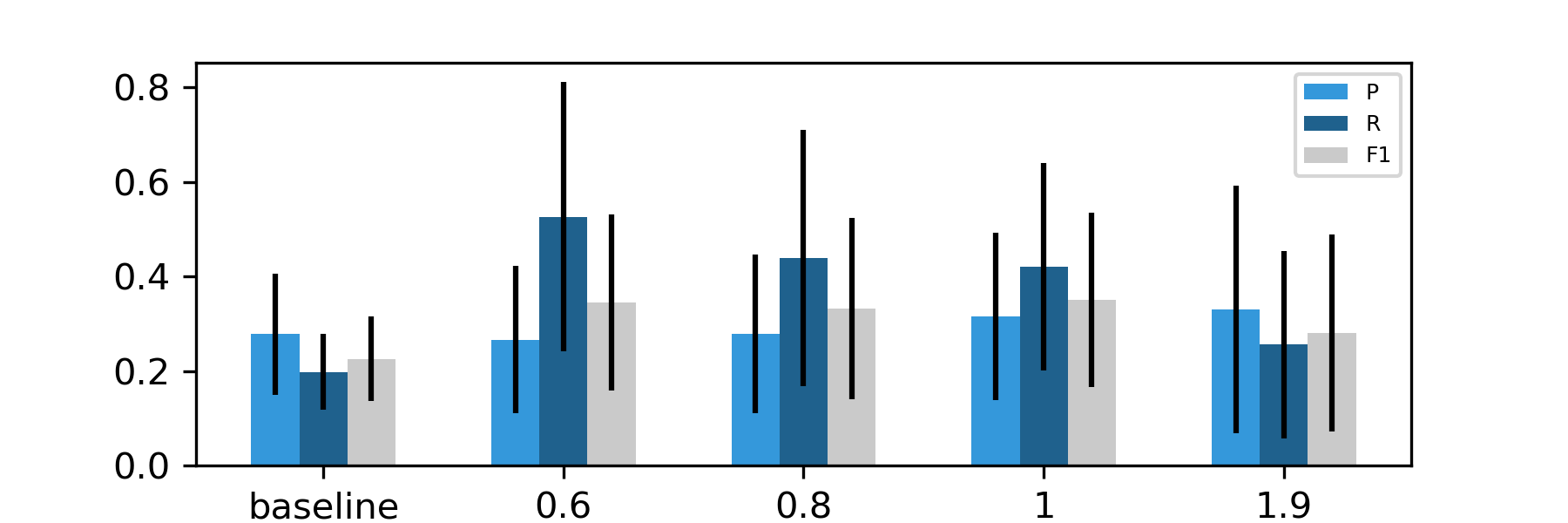}
}
\subfloat[G-PELT 1\_bar with $\alpha$ variation and $\beta$=0.15\label{1a}]{%
    \centering
    \includegraphics[width=\columnwidth]{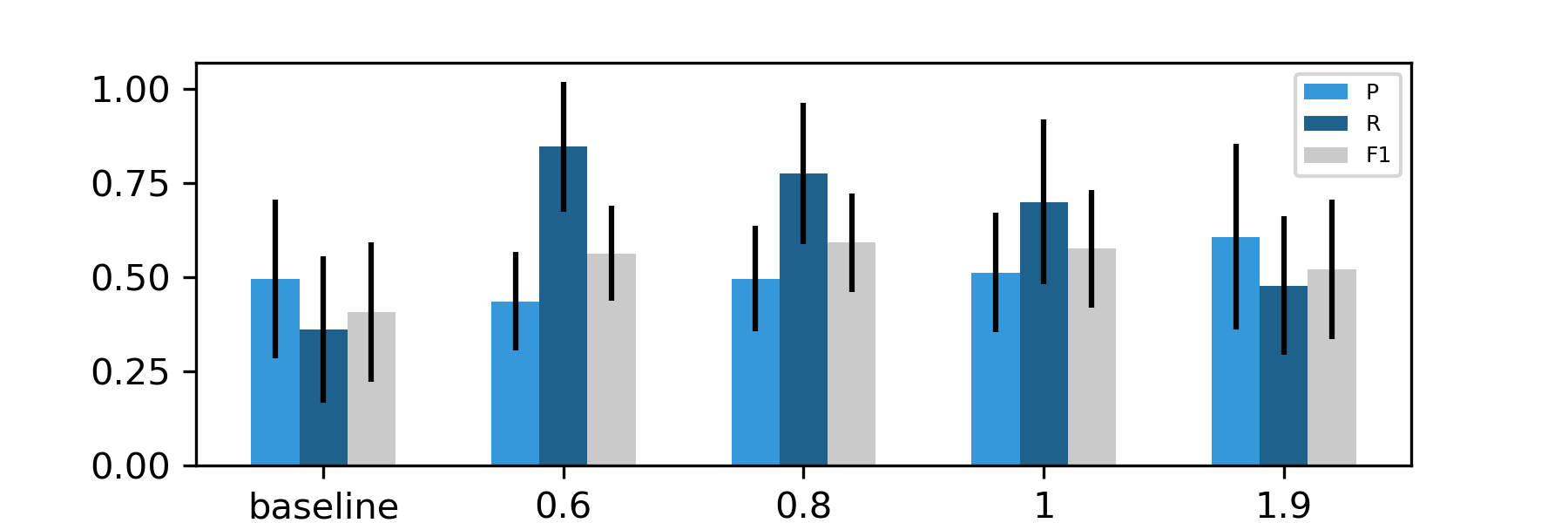}
} \\
\subfloat[G-PELT 1\_beat with $\beta$ variation and $\alpha$=0.6\label{1a}]{%
    \centering
    \includegraphics[width=\columnwidth]{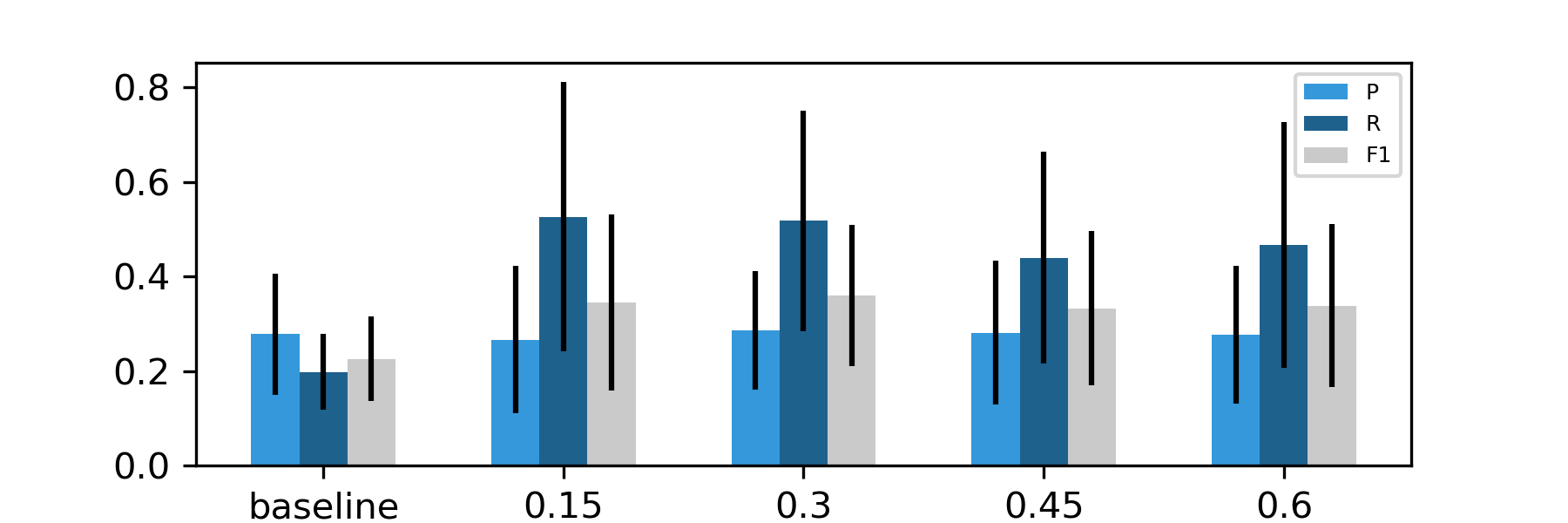}
}
\subfloat[G-PELT 1\_bar with $\beta$ variation and $\alpha$=0.6\label{1a}]{%
    \centering
    \includegraphics[width=\columnwidth]{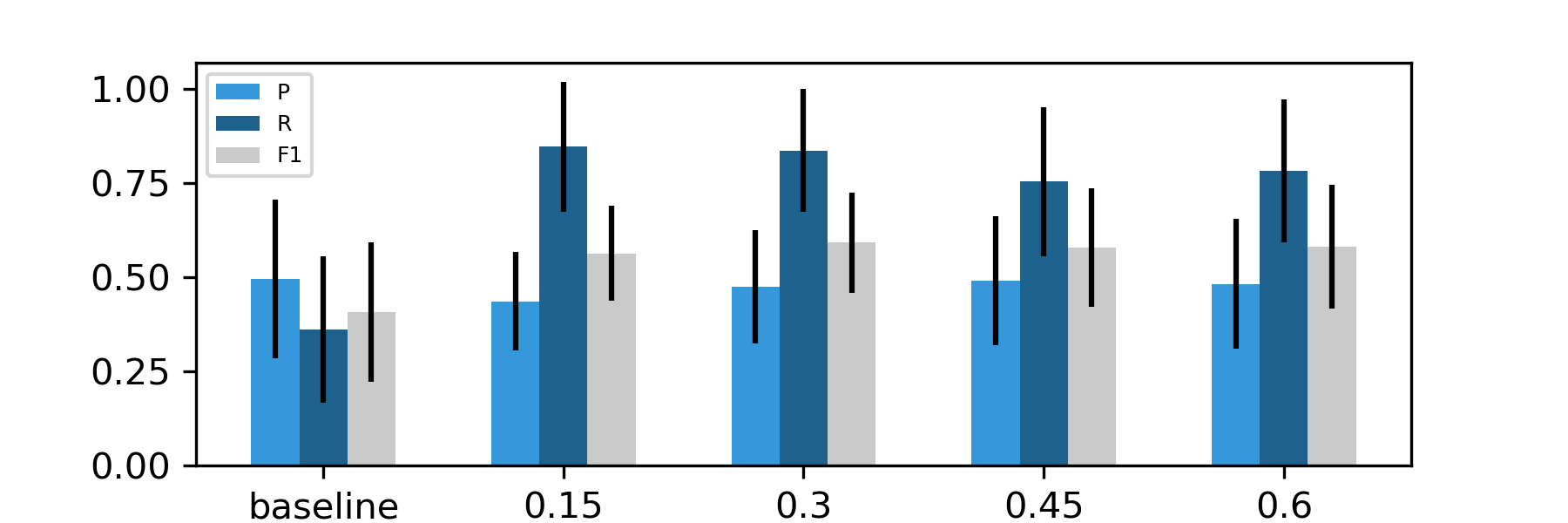}
} \\
\caption{Comparison of G-PELT method in the SWD dataset setting different $\alpha$ and $\beta$ values for 1 beat and 1 bar tolerances. The penalty value is fixed to 0.7.}
\label{fig:pel_min_size}
\end{figure*}

We follow the same procedure as done with the Norm method to find the optimal parameter values for the G-PELT method: $\alpha$ and $\beta$. We fix the penalty $p$ to 0.7 which we found to perform better than other values. In Fig. \ref{fig:pel_min_size} we show the results of our experiments with G-PELT method in the SWD.

Looking at the results in Fig. \ref{fig:pel_min_size}, we can see that the method is sensible to the $\alpha$ parameter, specially in the Recall. The best performing parameter values are $\alpha$=0.6, $\beta$=0.15 and $p$=0.7 for both 1 beat and 1 bar tolerances with a $F_1$=0.3455 for 1 beat and $F_1$=0.5640 for 1 bar tolerances (see Table \ref{tab:msa}).

\subsubsection{G-Window}
\begin{figure*}[!ht]
\subfloat[G-Window 1\_beat with $\alpha$ variation.\label{1a}]{%
    \centering
    \includegraphics[width=\columnwidth]{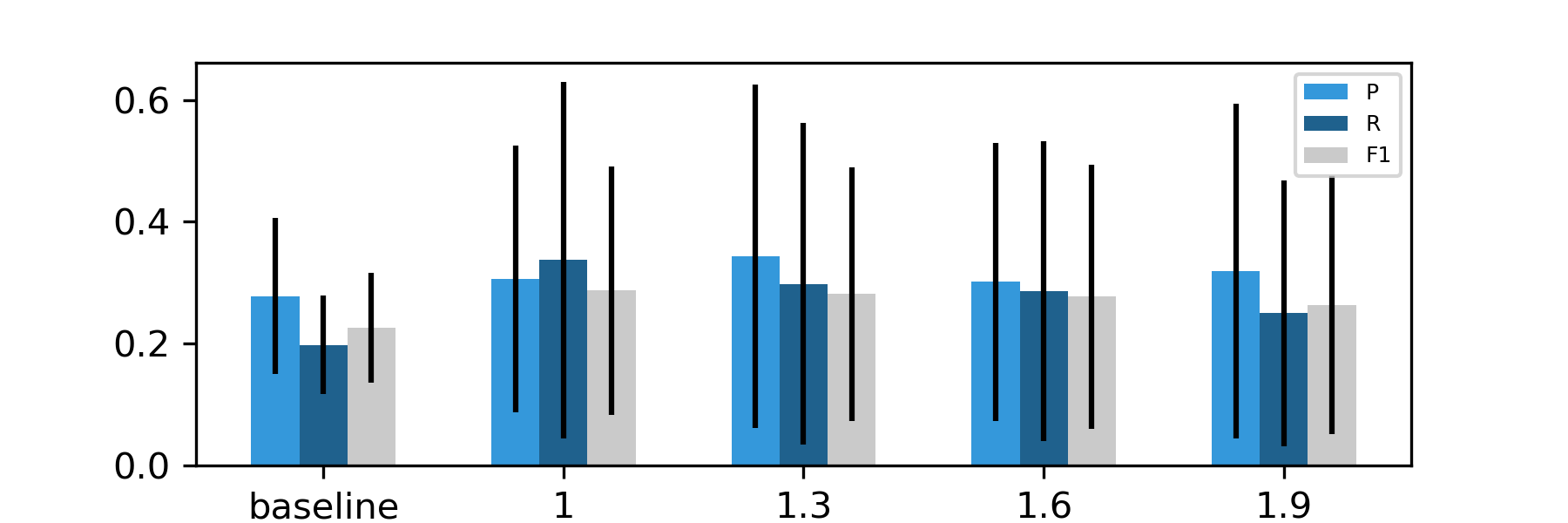}
}
\subfloat[G-Window 1\_bar with $\alpha$ variation.\label{1a}]{%
    \centering
    \includegraphics[width=\columnwidth]{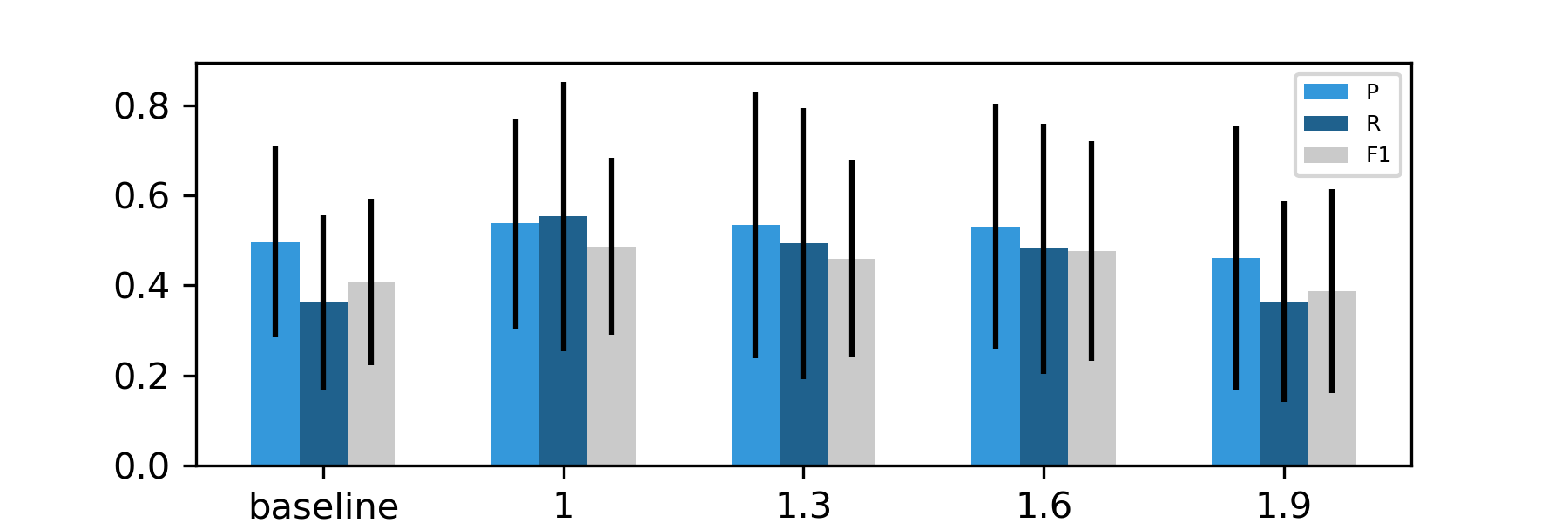}
} \\
\caption{Comparison of G-Window method in the SWD dataset setting different $\alpha$ values for \texttt{1\_beat} and \texttt{1\_bar} tolerances. The penalty value is fixed to 0.5.}
\label{fig:window_min_size}
\end{figure*}

As we did with the previous methods, we find the optimal parameter values for the G-Window method: $\alpha$ and $\beta$. We fix the penalty $p$ to 0.5 in this case. In Fig. \ref{fig:window_min_size} we show the results of our experiments with G-Window method in the SWD.

The results in Fig. \ref{fig:window_min_size} show that the G-Window method is also sensible to its parameter $\alpha$. We found that the best performing parameter values are $\alpha$=1 and $p$=0.5 for both 1 beat and 1 bar tolerances with a $F_1$=0.2869 for 1 beat and $F_1$=0.4863 (for 1 bar tolerances see Table \ref{tab:msa}).

\subsection{Beethoven sonatas for Piano (BPS)}
After testing the performance of our algorithms with the SWD, we now follow a similar procedure with the BPS dataset. In spite that the BPS dataset was originally proposed for symbolic harmonic analysis \cite{bps}, it also contains structure annotations of the first movements of Beethoven sonatas for piano at 3 levels, which makes it suitable for MSA \cite{GothamI19}. The structure annotations for each file are stored in \texttt{phrases.xlsx}\footnote{File 31 end has an error since a segment start must be minor than the end. We fixed it changing the value 246 to 346 that are the actual beats in the file.}. To test the performance of our methods with this dataset, we follow the same procedure that we described previously for the SWD. The only difference between this dataset and the SWD is that SWD provides only middle level annotations whereas in the BPS we will analyzed the low, middle and high structure levels since the dataset provides the annotations per level.
To show how the novelty of the graph adjacency matrix represents well the structure of the symbolic music, and to show an example of a file, we present 
in Fig. \ref{fig:msa-graph} the predicted boundaries in a sample of the BPS dataset with the G-PELT method. 

\subsubsection{G-PELT}
As we did for the SWD, we optimize the parameter values for the BPS dataset. However, since we found that the best performing method in the SWD was G-PELT, we will optimize the values of the parameters for this algorithm for the three structure levels and we will compare the performance in each level with the other methods (Norm and G-Window). In Table \ref{tab:msa_levels} we show the results of the ablation of the three methods for each structure level and tolerance.

\begin{figure*}[!ht]
\subfloat[High-level G-PELT 1\_bar with $\alpha$ variation and $\beta$=1.5.\label{1a}]{%
    \centering
    \includegraphics[width=\columnwidth]{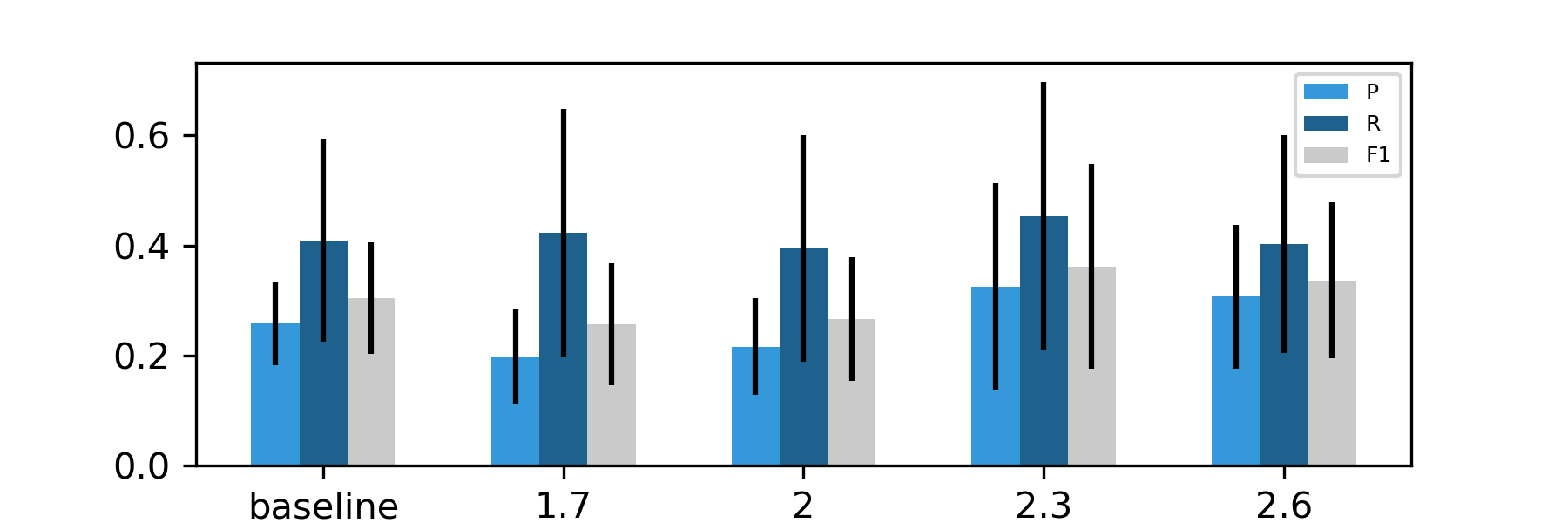}
}
\subfloat[High-level G-PELT 1\_bar with $\beta$ variation and $\alpha$=2.3.\label{1a}]{%
    \centering
\includegraphics[width=\columnwidth]{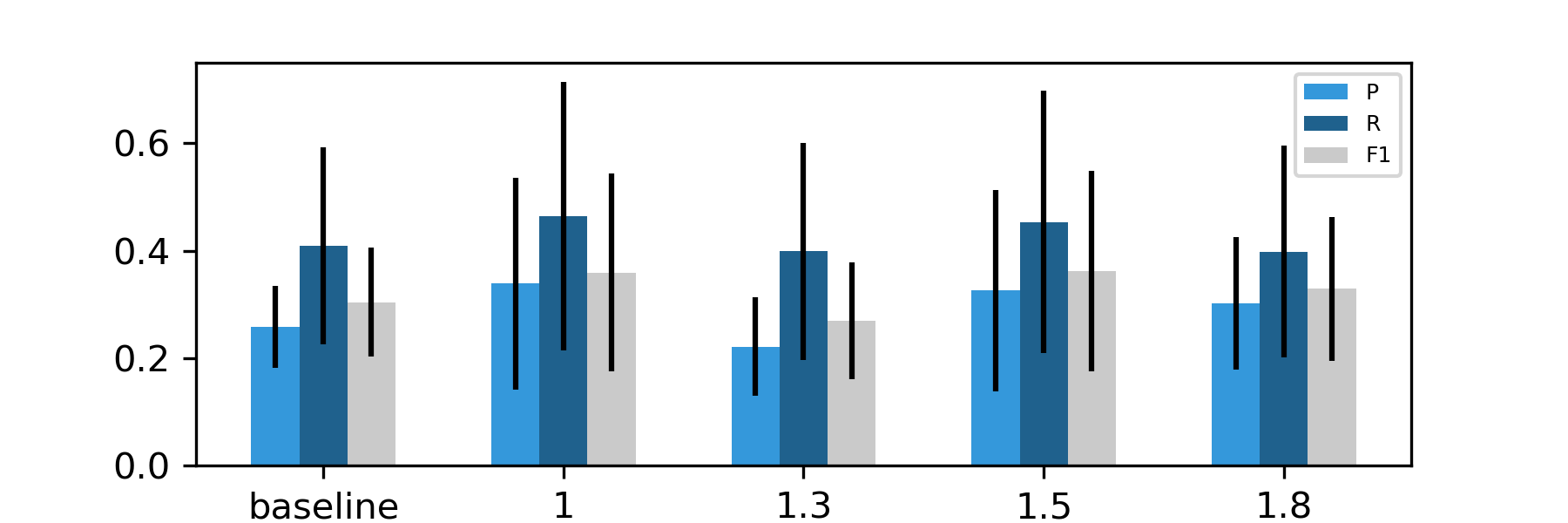}
} \\
\subfloat[Mid-level G-PELT 1\_beat with $\alpha$ variation and $\beta$=0.01.\label{1a}]{%
    \centering
    \includegraphics[width=\columnwidth]{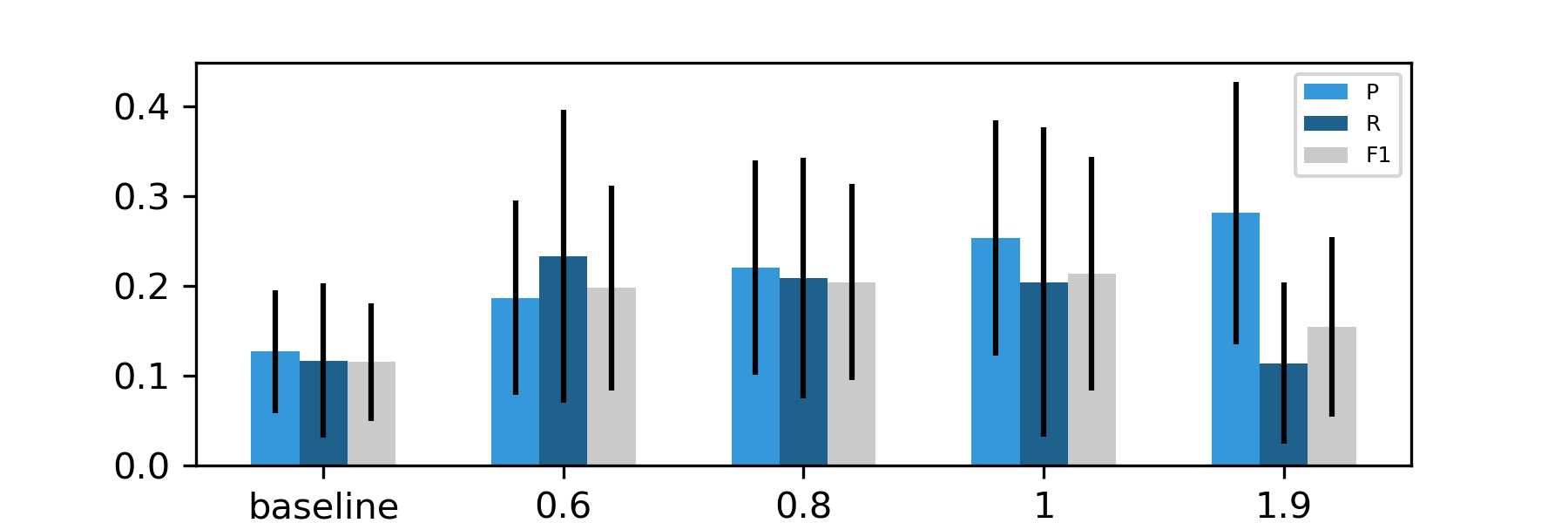}
}
\subfloat[Mid-level G-PELT 1\_beat with $\beta$ variation and $\alpha$=1.\label{1a}]{%
    \centering
    \includegraphics[width=\columnwidth]{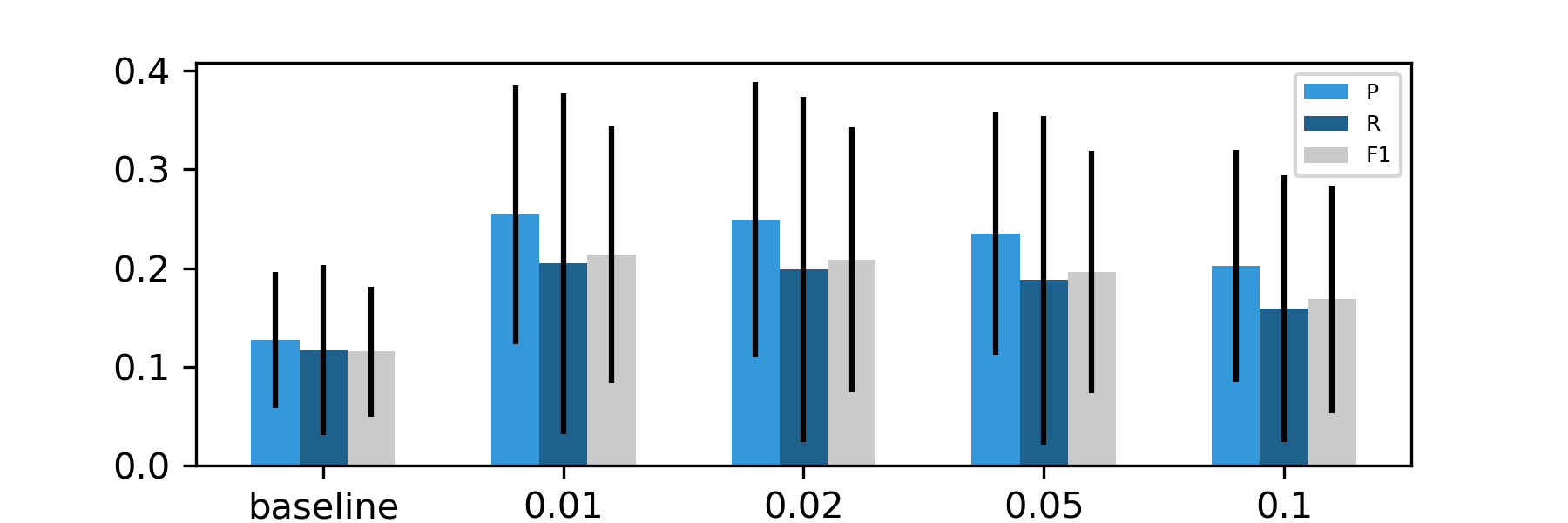}
} \\
\subfloat[Low-level G-PELT 1\_beat with $\alpha$ variation and $\beta$=0.15.\label{1a}]{%
    \centering
    \includegraphics[width=\columnwidth]{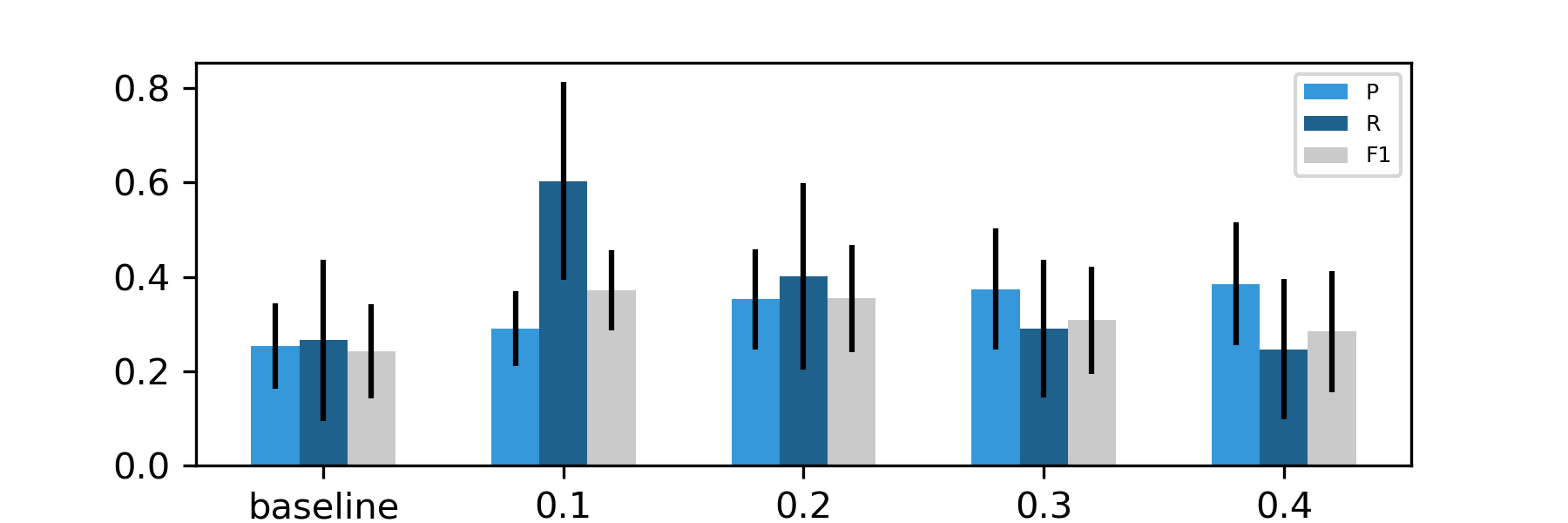}
}
\subfloat[Low-level G-PELT 1\_beat with $\beta$ variation and $\alpha$=0.1.\label{1a}]{%
    \centering
    \includegraphics[width=\columnwidth]{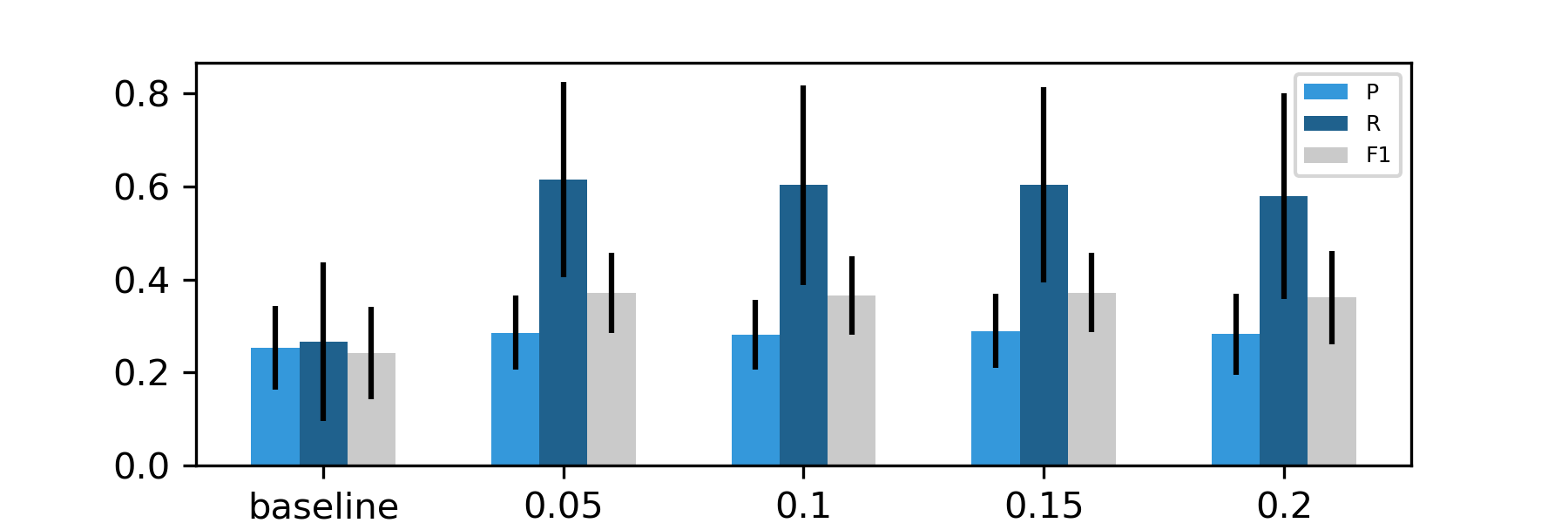}
} \\
\caption{Comparison of G-PELT method in the BPS dataset setting different $\alpha$ and $\beta$ values for the high, mid and low levels. High level is tested with 1 bar and mid and low levels with 1 beat tolerances. The penalty value is fixed to 4, 0.5 and 0.1 for the high, mid and low levels, respectively.}
\label{fig:pelt_bps}
\end{figure*}

Similar to the ablation with the SWD, we first vary $\alpha$ and then with the optimal $\alpha$ value we find the optimal $\beta$. The penalty is fixed and obtained for each structure level. In Fig. \ref{fig:pelt_bps} we show a comparison of different parameter values for G-PELT method in the structure levels (low, mid and high) in the BPS dataset.

Comparing the results with the SWD in general, we found similarities as for the high Recall values in comparison with the Precision. However, in the mid level the Precision is higher than the Recall for almost all the $\alpha$ and $\beta$ tested values. This means that the method predicts lower False Positives and higher False Negatives, or in other words, that the method will miss real boundaries (lower Recall) but the predicted boundaries will be more likely to be the real ones (higher Precision).

After the ablation of the parameters for each model and dataset, in Table \ref{tab:optimal} we show the optimal parameters for both SWD and BPS datasets.

\begin{table}[!h]
\caption{Optimal parameters and for each algorithm and dataset.}
    \centering
    \begin{tabular}{
    p{.5cm}
    >
    {\centering\arraybackslash}p{.5cm}
    >{\centering\arraybackslash}p{1.52cm}
    >{\centering\arraybackslash}p{4cm}
    }
    \toprule
         Dataset & level & Algorithm & Params\\
         \hline
        \multirow{4}{*}{SWD} & \multirow{4}{*}{mid} & \multirow{2}{*}{Norm} & $\alpha_{1}=0.6$, $\tau_{1}=1 $ \\
         & & & $\alpha_{2}=2$, $\tau_{2}=0.5$ \\
        & & G-PELT & $\alpha=0.6$, $\beta=0.15$, $p=0.7$ \\
        & & G-Window & $\alpha=1$, $p=0.5$ \\
         \cmidrule{1-4}
         \multirow{3}{*}{BPS} & \multirow{1}{*}{high} & G-PELT & $\alpha=2.3$, $\beta=1.5$, $p=4$ \\
        \cmidrule{2-4}
        & mid & G-PELT & $\alpha=1$, $\beta=0.01$, $p=0.5$ \\
        \cmidrule{2-4}
        & low & G-PELT & $\alpha=0.1$, $\beta=0.15$, $p=0.1$ \\
        \bottomrule
    \end{tabular}
    \label{tab:optimal}
\end{table}

\section{Evaluation} \label{sec:eval}
After finding the optimal parameter values of the algorithms in Section \ref{sec:ablation}, we test the model in the two datasets with the best hyperparameters. We also make an analysis of the errors in beats that the algorithm commits when it makes a prediction to proof its precision.

\begin{figure}
    \centering
    \includegraphics[width=\columnwidth]{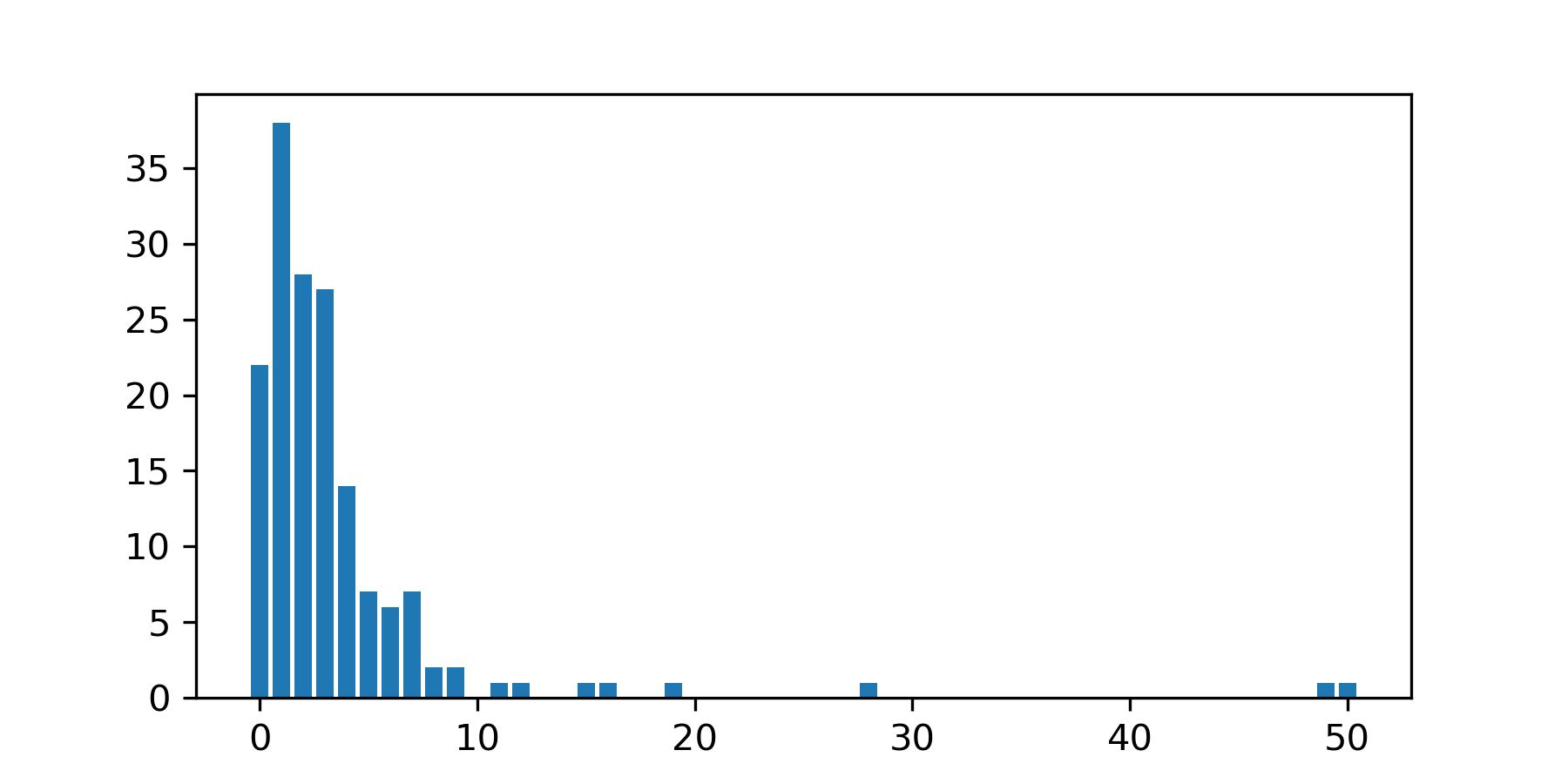}
    \caption{Number of errors (vertical axis) in beat units (horizontal axis) measured in the predicted boundaries with G-PELT method in the SWD dataset.}
    \label{fig:errors}
\end{figure}

We measured the errors across the datasets to give more insights about the algorithms precision in terms of time tolerance. In Fig. \ref{fig:errors} we show the histogram of errors in beats that the G-PELT algorithm makes in relation with the number of boundaries detected.
We can observe that the most common error rage is from 0 to 10 beats, which means that when the algorithm makes a prediction, it is relatively close to the real boundary.

The results in Table \ref{tab:msa} show  the performance in the SWD, however, the analysis can be confusing due to the fact that the SWD does not have annotations per structure level. In order to do a deeper analysis and show how the algorithms perform in the different levels of the structure of music, we also test the algorithms with the BPS dataset. This dataset provides annotations at the 3 structure levels in Western classical music. Since the dataset contains the Beethoven piano sonatas, the levels are organized as: high-level structure (exposition, development and recapitulation), themes (1st, tr, 2nd, coda, etc), and phrases or motifs (a, a', etc). Low level boundaries contain the mid and high level ones, and mid level boundaries also contain high level ones. With this annotations we can measure the algorithms performance in various levels and demonstrate in which level they perform better.
In Table \ref{tab:msa_levels} we show the performance metrics of the algorithms measured with the BPS dataset.

\begin{table*}[!t]
    \centering
    \caption{MSA results for the algorithms with optimal parameter values measured with the BPS dataset.}
    \begin{tabular}{
    p{1.1cm}|  
    >{\centering\arraybackslash}p{1.2cm}
    >{\centering\arraybackslash}p{1.2cm}
    >{\centering\arraybackslash}p{1.5cm}|
    >{\centering\arraybackslash}p{1.2cm}
    >{\centering\arraybackslash}p{1.2cm}
    >{\centering\arraybackslash}p{1.5cm}|
    >{\centering\arraybackslash}p{1.2cm}
    >{\centering\arraybackslash}p{1.2cm}
    >{\centering\arraybackslash}p{1.15cm}
    }
    \cmidrule{1-10}
    & \multicolumn{3}{c|}{High Level} & \multicolumn{3}{c|}{Mid Level} & \multicolumn{3}{c|}{Low Level}\\
    \hline
    & P & R & $F_1$ & P & R & $F_1$ & P & R & $F_1$\\
    \cmidrule{1-10}
        \rowcolor{Gray}
        & \multicolumn{9}{c|}{Tolerance: 1 beat}\\
        \cmidrule{1-10}
        baseline & $0.1875_{\pm 0.10}$ & $0.3041_{\pm 0.21}$ & $0.2233_{\pm 0.13}$ & $0.1272_{\pm 0.06}$ & $0.1169_{\pm 0.08}$ & $0.1155_{\pm 0.06}$ & $0.2533_{\pm 0.09}$ & $0.2656_{\pm 0.17}$ & $0.2417_{\pm 0.09}$ \\
        Norm $\mathbf{b}$ & $0.0579_{\pm 0.03}$ & $0.4559_{\pm 0.23}$ & $0.0987_{\pm 0.05}$ &  $0.1036_{\pm 0.04}$ & $0.3773_{\pm 0.23}$ & 
        $0.1573_{\pm 0.07}$ &
        $0.1829_{\pm 0.06}$ & $0.7612_{\pm 0.19}$ & $0.2857_{\pm 0.08}$ \\
        Norm $\mathbf{b}^\prime$  & $0.2585_{\pm 0.13}$ & $0.3906_{\pm 0.15}$ & $\mathbf{0.2880_{\pm 0.10}}$ &  $0.2378_{\pm 0.14}$ & $0.1038_{\pm 0.07}$  & $0.1308_{\pm 0.07}$ & $0.2969_{\pm 0.12}$ & $0.2450_{\pm 0.17}$ & $0.2489_{\pm 0.13}$\\
        G-PELT & $0.2765_{\pm 0.16}$ & $0.3114_{\pm 0.20}$ & $0.2686_{\pm 0.15}$ & $0.2540_{\pm 0.13}$ & $0.2045_{\pm 0.17}$ & $\mathbf{0.2137_{\pm 0.13}}$ & $0.2896_{\pm 0.07}$ & $0.6033_{\pm 0.20}$ & $\mathbf{0.3716_{\pm 0.08}}$\\
        G-Wind. & $0.2092_{\pm 0.25}$ & $0.2614_{\pm 0.31}$ & $0.2183_{\pm 0.26}$ & $0.2436_{\pm 0.15}$ & $0.1531_{\pm 0.11}$ & $0.1724_{\pm 0.10}$ & $0.4816_{\pm 0.22}$ & $0.1205_{\pm 0.08}$ & $0.1849_{\pm 0.11}$ \\
        \hline
        \rowcolor{Gray}
        & \multicolumn{9}{c|}{Tolerance: 1 bar}\\
        \hline
        baseline & $0.2578_{\pm 0.07}$ & $0.4088_{\pm 0.18}$ & $0.3040_{\pm 0.10}$ & $0.2611_{\pm 0.08}$ & $0.2557_{\pm 0.17}$ & $0.2447_{\pm 0.11}$ & $0.4415_{\pm 0.08}$ & $0.4676_{\pm 0.25}$ & $0.4254_{\pm 0.13}$ \\
        Norm $\mathbf{b}$ & $0.0862_{\pm 0.06}$ & $0.6432_{\pm 0.26}$ & $0.1455_{\pm 0.09}$ & $0.1820_{\pm 0.06}$ & $0.6347_{\pm 0.25}$ & $0.2755_{\pm 0.09}$ & $0.2506_{\pm 0.06}$ & $0.9711_{\pm 0.19}$ & $0.3854_{\pm 0.08}$ \\
        Norm $\mathbf{b}^\prime$ & $0.2708_{\pm 0.13}$ & $0.4140_{\pm 0.18}$ & $0.3028_{\pm 0.11}$ & $0.3674_{\pm 0.20}$ & $0.1721_{\pm 0.13}$ & $0.2132_{\pm 0.13}$ & $0.5904_{\pm 0.12}$ & $0.4458_{\pm 0.17}$ & $0.4696_{\pm 0.13}$ \\
        G-PELT & $0.3437_{\pm 0.27}$ & $0.3585_{\pm 0.30}$ & $\mathbf{0.3361_{\pm 0.26}}$ & $0.3865_{\pm 0.17}$ & $0.3024_{\pm 0.19}$ & $\mathbf{0.3224_{\pm 0.16}}$ & $0.4370_{\pm 0.15}$ & $0.8434_{\pm 0.13}$ & $\mathbf{0.5473_{\pm 0.12}}$ \\
        G-Wind. & $0.2466_{\pm 0.27}$ & $0.3192_{\pm 0.36}$ & $0.2617_{\pm 0.28}$ & $0.3566_{\pm 0.19}$ & $0.2345_{\pm 0.16}$ & $0.2611_{\pm 0.14}$ & $0.7538_{\pm 0.17}$ & $0.1989_{\pm 0.12}$ & $0.2999_{\pm 0.15}$ \\
         \bottomrule
    \end{tabular}
    \label{tab:msa_levels}
\end{table*}

\begin{figure*}
    \centering
    \includegraphics[width=\textwidth]{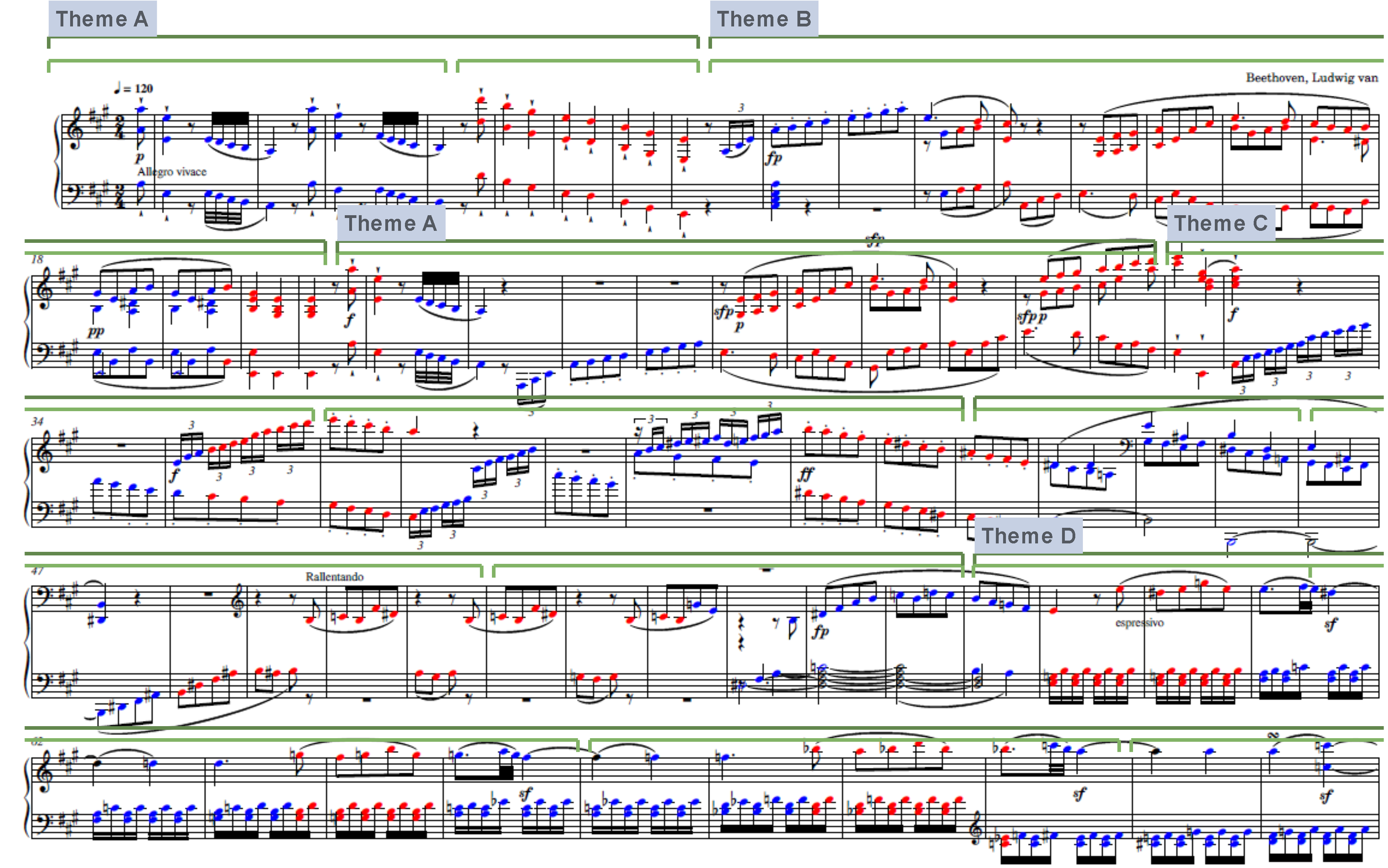}
    \caption{Beginning of the 1st movement of the Beethoven Piano Sonata n2 segmented by our G-PELT method with optimal parameter values. Predicted segments are colored in red and blue. Ground truth is marked in bright green (low level, 5th column in the csv) and dark green (mid level, 6th column in the csv) lines above the staff.}
    \label{fig:test}
\end{figure*}

In \ref{ap:2} we see another example of the G-PELT algorithm performance in a Debussy piece, which differs in period and form from the SWD for wich the parameters have beeen optimized.

\section{Discussion} \label{sec:discussion}
The best-performing algorithm for the SWD is G-PELT as we show in Table \ref{tab:msa} with $P=0.4366$, $R=0.8473$ and $F_1=0.5640$ (optimal parameters are shown in Table \ref{tab:optimal}). The high Recall value means that the algorithm is capable of identify boundaries in the ground truth and that it does not predict boundaries that are not in the ground truth (low number of False Negatives). However, the low Precision shows that the algorithm does not detect a high amount of the real boundaries (high number of False Positives). This is an indicator that the algorithm could be used in scenarios where the recall of the algorithm is a priority.

When testing the performance of the methods in the BPS dataset, we find differences in the results depending on the structure level we are working with. About the performance in the high-level, we should clarify that it is more difficult to predict only 3 to 5 boundaries versus the 40 to 60 that the low-level presents per file (see Table \ref{tab:data}). However, the algorithms detect the end of the file boundary (that have been added also in the baselines), which is translated as the addition of a True Positive. The impact of this in the high-level, that has less boundaries per file, leads to having a higher impact than in the mid and low levels. The reason why the expectation of a decrease in the performance in the high-level analysis come from the fact that if a file has $\sim$800 beats (BPS file n1) and 3 boundaries in the high level, the boundaries represent less than the 0.5\% of the beats making it difficult for the algorithm to perform more accurately at lower tolerance levels (\texttt{1\_beat}). It is worth noting that the same applies to the baseline, as the number of boundaries in annotations increases, it becomes more likely to increase the number of true positives because there are also more synthetic boundaries per file.

An example of the performance of the G-PELT method in a Beethoven piano sonata is shown in Fig. \ref{fig:test}. In the figure, we show the predicted sections with colors and the ground truth annotations of the BPS dataset in the green lines above the score\footnote{The ground truth has been extracted from \url{https://github.com/MarkGotham/Taking-Form/blob/master/corpus/Beethoven_Sonatas/sonata2op2no2movt1.csv}}. We would like to highlight how the algorithm identifies sections that are not labeled in the dataset annotations, such as cadences. As an example, there is an annotation in bar 58 (theme D start) that, for our understanding, does not correspond to the real boundary. In this case, the method identifies as the real boundary the bar 59 which is more likely to be the real one due to the rhythm changes in the left hand. The same happens in bar 32 (theme C) and in bar 66 where the annotation seem to be displaced from the real boundary. Being that said, we would suggest that the BPS dataset structure annotations should be revised.

\section{Conclusion} \label{sec:conclusions}

We presented three methods that aim to segment symbolic music in its structure or form. After measuring the performance of each method in the context of symbolic structure segmentation with two public datasets, we can conclude that the best performing method is G-PELT for both SWD and BPS datasets and structure levels.
We give evidence of how by changing the parameters of the algorithms, music can be segmented into different levels of structure, opening new possibilities to better understand how music is made and opening new scenarios in applications such as music generation or classification.
The proposed methods are online and unsupervised which makes them usable to process large corpus of data, however, for a deeper analysis of a particular form, a learning algorithm might be needed to improve the results of this work.
We suggest to revisit the annotations of the BPS dataset by a group of expert musicians. This could be done with the help of our G-PELT method which is our best performing method, as a guide when analyzing the low level structure.

\subsection{Applications}
The presented segmentation method can be used in applications such as music generation or data augmentation. In music generation, current tokenizers such as the Multi-Track Music Machine (MMM) \cite{ens2020mmm} do not provide section tokens due to the fact that they are not encoded in MIDI-like nor score-like formats. With our method, these section tokens could be included (at beat or bar level) to generate symbolic music and being able to inpaint sections, e.g., readapting the MMMBar tokenization to include section tokens. The fast online proposed method helps tokenizing the files faster than offline methods.

Apart from that, in applications such as fingering, data augmentation is difficult due to the fact that common techniques (pich shifting, etc) cannot be applied since they change the fingering or difficulty of the task. Being able to segment music in coherent parts might help these models for training purposes. Automatic structural segmentation can also help annotators when it comes to annotating the structure at a large corpus of data.

Future lines of work might be building a deep neural graph network that segments music based on this methods. In addition, having a better performance in the segmentation task might open future works about structure labelling or classification. Apart from that, since we only use temporal information to construct the graph, we propose adding harmonic information in the novelty curve obtained from the adjacency matrix to improve the results of this work. We did not include them due to the fact that these features need to be predicted as they are not encoded in a MIDI file, but a future work might be to encode the chord progression predicted with recent work on harmonic analysis (e.g. AugmentedNet \cite{augmentednet}) to better capture the structure boundaries since harmony and structure are closely related to each other.





\appendix
\section{Structure Encoding Example}
\begin{figure*}[!h]
 \centerline{
 \includegraphics[width=.7\textwidth]{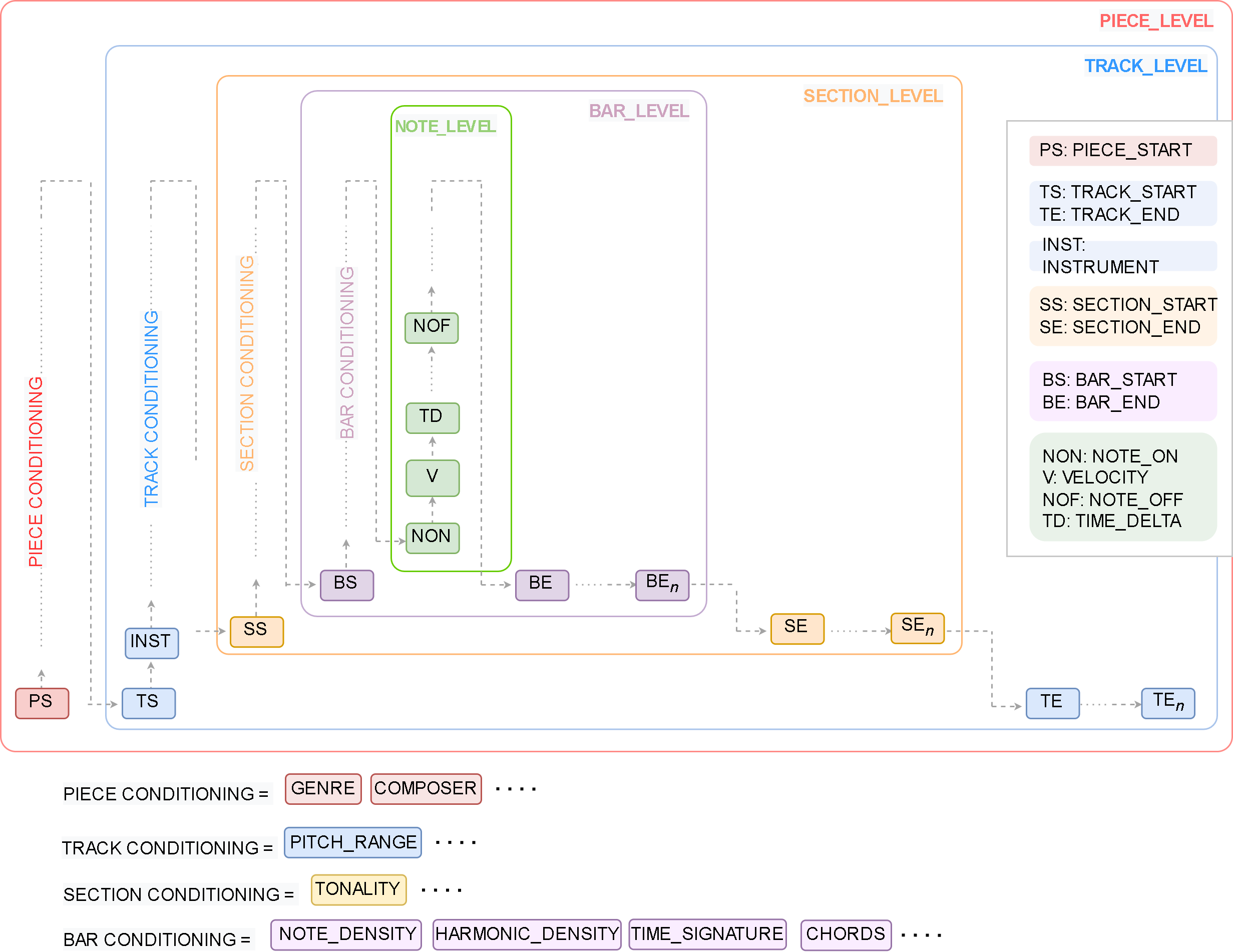}}
 \caption{A general encoding based on the MMM. The figure shows a general scheme configuration of how the MMMTrack encoding extended for conditioning the generation at different levels. We extend the three levels proposed in the original MMM by adding a structure level.}
 \label{fig:tokens}
\end{figure*}

Sctructure segmentation can be used to tokenize sections and train large language models that generate music. In Fig. \ref{fig:tokens} we show an example based on the MMMTrack tokenizer \cite{ens2020mmm} that includes the section tokens. 

\section{G-PELT Test Example} \label{ap:2}

With the G-PELT algorithm with optimized parameters with SWD we show how the algorithm works in other music styles and periods. In Fig. \ref{fig:test} we show an example of a Vals by Debussy which is a different form and a composer from a different period in the Western classical music than the Schubert's Winterreise songs.
In the figure, it can be seen that the algorithm segments coherent sections attending to the rhtyhm and texture of the piece, in spite that the parameter values have been optimized for other form and style.

\begin{figure*}
    \centering
    \includegraphics[width=\textwidth]{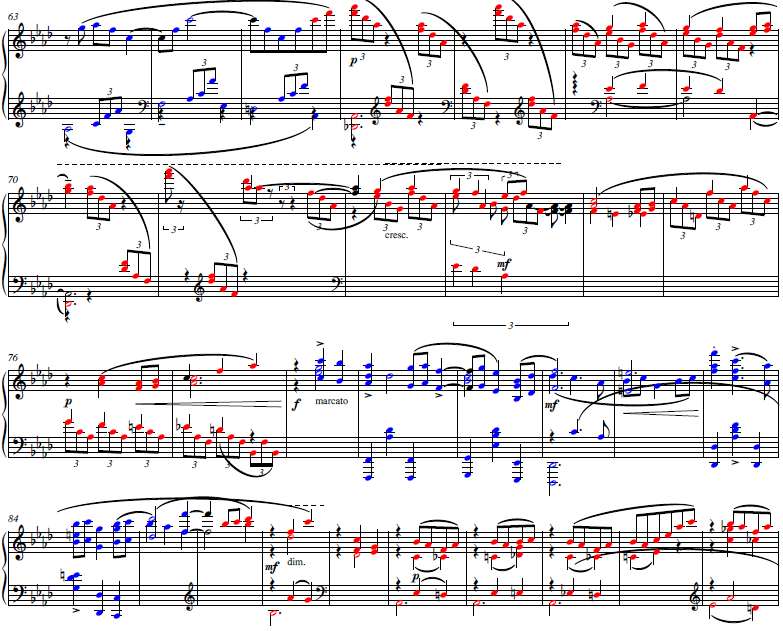}

\caption{Vals romantique by Debussy (bars 63-93) segmented by our G-PELT method with parameters optimized for Shubert Winterreise Dataset.}
\label{fig:test}
\end{figure*}

\section*{Acknowledgment}

The authors would like to thank Pedro Ramoneda for his insightful comments.

This research has been partially supported by the Spanish Science, Innovation and University Ministry by the RTI2018-096986-B-C31 contract and the Aragonese Government by the AffectiveLab-T60-20R project.

\ifCLASSOPTIONcaptionsoff
  \newpage
\fi



%

%

\bibliographystyle{IEEEtran}
\bibliography{references}

\begin{thebibliography}{10}
\providecommand{\url}[1]{#1}
\csname url@samestyle\endcsname
\providecommand{\newblock}{\relax}
\providecommand{\bibinfo}[2]{#2}
\providecommand{\BIBentrySTDinterwordspacing}{\spaceskip=0pt\relax}
\providecommand{\BIBentryALTinterwordstretchfactor}{4}
\providecommand{\BIBentryALTinterwordspacing}{\spaceskip=\fontdimen2\font plus
\BIBentryALTinterwordstretchfactor\fontdimen3\font minus
  \fontdimen4\font\relax}
\providecommand{\BIBforeignlanguage}[2]{{%
\expandafter\ifx\csname l@#1\endcsname\relax
\typeout{** WARNING: IEEEtran.bst: No hyphenation pattern has been}%
\typeout{** loaded for the language `#1'. Using the pattern for}%
\typeout{** the default language instead.}%
\else
\language=\csname l@#1\endcsname
\fi
#2}}
\providecommand{\BIBdecl}{\relax}
\BIBdecl

\bibitem{review}
\BIBentryALTinterwordspacing
C.~Hernandez-Olivan and J.~R. Beltr{\'a}n, \emph{Music Composition with Deep
  Learning: A Review}.\hskip 1em plus 0.5em minus 0.4em\relax Cham: Springer
  International Publishing, 2023, pp. 25--50. [Online]. Available:
  \url{https://doi.org/10.1007/978-3-031-18444-4_2}
\BIBentrySTDinterwordspacing

\bibitem{lerdahl1983overview}
F.~Lerdahl and R.~Jackendoff, ``An overview of hierarchical structure in
  music,'' \emph{Music Perception}, pp. 229--252, 1983.

\bibitem{lopez2012automatic}
M.~R. L{\'o}pez and A.~Volk, ``Automatic segmentation of symbolic music
  encodings: A survey,'' \emph{Utrecht University}, 2012.

\bibitem{rodriguez2014comparing}
M.~Rodr{\'\i}guez~L{\'o}pez, B.~de~Haas, A.~Volk \emph{et~al.}, ``Comparing
  repetition-based melody segmentation models,'' in \emph{Proceedings of the
  9th Conference on Interdisciplinary Musicology (CIM14)}.\hskip 1em plus 0.5em
  minus 0.4em\relax SIMPK and ICCMR, 2014, pp. 143--148.

\bibitem{lattner2015probabilistic}
S.~Lattner, M.~Grachten, K.~Agres, and C.~E. Cancino~Chac{\'o}n,
  ``Probabilistic segmentation of musical sequences using restricted boltzmann
  machines,'' in \emph{International Conference on Mathematics and Computation
  in Music}.\hskip 1em plus 0.5em minus 0.4em\relax Springer, 2015, pp.
  323--334.

\bibitem{bassan2022unsupervised}
S.~Bassan, Y.~Adi, and J.~S. Rosenschein, ``Unsupervised symbolic music
  segmentation using ensemble temporal prediction errors,'' \emph{arXiv
  preprint arXiv:2207.00760}, 2022.

\bibitem{arxiv.2205.08579}
G.~Wu, S.~Liu, and X.~Fan, ``The power of fragmentation: A hierarchical
  transformer model for structural segmentation in symbolic music generation,''
  2022.

\bibitem{msareview}
O.~Nieto, G.~J. Mysore, C.~Wang, J.~B.~L. Smith, J.~Schl{\"{u}}ter, T.~Grill,
  and B.~McFee, ``Audio-based music structure analysis: Current trends, open
  challenges, and applications,'' \emph{Transactions of the International
  Society for Music Information Retrieval}, vol.~3, no.~1, pp. 246--263, 2020.

\bibitem{muller2015}
\BIBentryALTinterwordspacing
M.~M{\"u}ller, \emph{Music Structure Analysis}.\hskip 1em plus 0.5em minus
  0.4em\relax Cham: Springer International Publishing, 2015, pp. 167--236.
  [Online]. Available: \url{https://doi.org/10.1007/978-3-319-21945-5_4}
\BIBentrySTDinterwordspacing

\bibitem{foote1999visualizing}
J.~Foote, ``Visualizing music and audio using self-similarity,'' in
  \emph{Proceedings of the 7th {ACM} International Conference on Multimedia
  '99, Orlando, FL, USA, October 30 - November 5, 1999, Part 1}, J.~F. Buford,
  S.~M. Stevens, D.~C.~A. Bulterman, K.~Jeffay, and H.~Zhang, Eds.\hskip 1em
  plus 0.5em minus 0.4em\relax {ACM}, 1999, pp. 77--80.

\bibitem{boundaries2021hernandezolivan}
C.~Hernandez-Olivan, J.~R. Beltran, and D.~Diaz-Guerra, ``Music boundary
  detection using convolutional neural networks: A comparative analysis of
  combined input features,'' \emph{International Journal of Interactive
  Multimedia and Artificial Intelligence}, vol.~7, no.~2, pp. 78--88, 2021.

\bibitem{serra2012unsupervised}
J.~Serr{\`{a}}, M.~M{\"{u}}ller, P.~Grosche, and J.~L. Arcos, ``Unsupervised
  detection of music boundaries by time series structure features,'' in
  \emph{Proceedings of the Twenty-Sixth {AAAI} Conference on Artificial
  Intelligence, July 22-26, 2012, Toronto, Ontario, Canada}, J.~Hoffmann and
  B.~Selman, Eds.\hskip 1em plus 0.5em minus 0.4em\relax {AAAI} Press, 2012.

\bibitem{PanagakisKA11}
\BIBentryALTinterwordspacing
Y.~Panagakis, C.~Kotropoulos, and G.~R. Arce, ``l1-graph based music structure
  analysis,'' in \emph{Proceedings of the 12th International Society for Music
  Information Retrieval Conference, {ISMIR} 2011, Miami, Florida, USA, October
  24-28, 2011}, A.~Klapuri and C.~Leider, Eds.\hskip 1em plus 0.5em minus
  0.4em\relax University of Miami, 2011, pp. 495--500. [Online]. Available:
  \url{http://ismir2011.ismir.net/papers/PS4-4.pdf}
\BIBentrySTDinterwordspacing

\bibitem{meredith2002algorithms}
D.~Meredith, K.~Lemstr{\"o}m, and G.~A. Wiggins, ``Algorithms for discovering
  repeated patterns in multidimensional representations of polyphonic music,''
  \emph{Journal of New Music Research}, vol.~31, no.~4, pp. 321--345, 2002.

\bibitem{lerdahl1996generative}
F.~Lerdahl and R.~S. Jackendoff, \emph{A Generative Theory of Tonal Music,
  reissue, with a new preface}.\hskip 1em plus 0.5em minus 0.4em\relax MIT
  press, 1996.

\bibitem{Cambouropoulos01}
\BIBentryALTinterwordspacing
E.~Cambouropoulos, ``The local boundary detection model {(LBDM)} and its
  application in the study of expressive timing,'' in \emph{Proceedings of the
  2001 International Computer Music Conference, {ICMC} 2001, Havana, Cuba,
  September 17-22, 2001}.\hskip 1em plus 0.5em minus 0.4em\relax Michigan
  Publishing, 2001. [Online]. Available:
  \url{https://hdl.handle.net/2027/spo.bbp2372.2001.021}
\BIBentrySTDinterwordspacing

\bibitem{temperley2004cognition}
D.~Temperley, \emph{The cognition of basic musical structures}.\hskip 1em plus
  0.5em minus 0.4em\relax MIT press, 2004.

\bibitem{cenkerova2018crossing}
Z.~Cenkerov{\'a}, M.~Hartmann, and P.~Toiviainen, ``Crossing phrase boundaries
  in music,'' in \emph{Proceedings of the Sound and Music Computing
  Conferences}, 2018.

\bibitem{cambouropoulos2004influence}
E.~Cambouropoulos and C.~Tsougras, ``Influence of musical similarity on melodic
  segmentation: Representations and algorithms,'' in \emph{Journ{\'e}es
  d'informatique musicale}, 2004.

\bibitem{wilder2008adaptive}
G.~Wilder, ``Adaptive melodic segmentation and motivic identification,'' in
  \emph{ICMC}, 2008.

\bibitem{Kranenburg20}
\BIBentryALTinterwordspacing
P.~van Kranenburg, ``Rule mining for local boundary detection in melodies,'' in
  \emph{Proceedings of the 21th International Society for Music Information
  Retrieval Conference, {ISMIR} 2020, Montreal, Canada, October 11-16, 2020},
  J.~Cumming, J.~H. Lee, B.~McFee, M.~Schedl, J.~Devaney, C.~McKay,
  E.~Zangerle, and T.~de~Reuse, Eds., 2020, pp. 271--278. [Online]. Available:
  \url{http://archives.ismir.net/ismir2020/paper/000226.pdf}
\BIBentrySTDinterwordspacing

\bibitem{LattnerCG15}
S.~Lattner, C.~E.~C. Chac{\'{o}}n, and M.~Grachten, ``Pseudo-supervised
  training improves unsupervised melody segmentation,'' in \emph{Proceedings of
  the Twenty-Fourth International Joint Conference on Artificial Intelligence,
  {IJCAI} 2015, Buenos Aires, Argentina, July 25-31, 2015}, Q.~Yang and M.~J.
  Wooldridge, Eds.\hskip 1em plus 0.5em minus 0.4em\relax {AAAI} Press, 2015,
  pp. 2459--2465.

\bibitem{LopezV13}
M.~E. Rodr{\'{\i}}guez{-}L{\'{o}}pez and A.~Volk, ``Symbolic segmentation: {A}
  corpus-based analysis of melodic phrases,'' in \emph{Sound, Music, and Motion
  - 10th International Symposium, {CMMR} 2013, Marseille, France, October
  15-18, 2013. Revised Selected Papers}, ser. Lecture Notes in Computer
  Science, M.~Aramaki, O.~Derrien, R.~Kronland{-}Martinet, and S.~Ystad, Eds.,
  vol. 8905.\hskip 1em plus 0.5em minus 0.4em\relax Springer, 2013, pp.
  548--557.

\bibitem{tavern}
J.~Devaney, C.~Arthur, N.~Condit{-}Schultz, and K.~Nisula, ``Theme and
  variation encodings with roman numerals {(TAVERN):} {A} new data set for
  symbolic music analysis,'' in \emph{Proceedings of the 16th International
  Society for Music Information Retrieval Conference, {ISMIR} 2015,
  M{\'{a}}laga, Spain, October 26-30, 2015}, M.~M{\"{u}}ller and F.~Wiering,
  Eds., 2015, pp. 728--734.

\bibitem{10.2307/43829264}
\BIBentryALTinterwordspacing
M.~Giraud, R.~Groult, E.~Leguy, and F.~Levé, ``Computational fugue analysis,''
  \emph{Computer Music Journal}, vol.~39, no.~2, pp. 77--96, 2015. [Online].
  Available: \url{http://www.jstor.org/stable/43829264}
\BIBentrySTDinterwordspacing

\bibitem{allegraud2019learning}
P.~Allegraud, L.~Bigo, L.~Feisthauer, M.~Giraud, R.~Groult, E.~Leguy, and
  F.~Lev{\'e}, ``Learning sonata form structure on mozart's string quartets,''
  \emph{Transactions of the International Society for Music Information
  Retrieval (TISMIR)}, vol.~2, no.~1, pp. 82--96, 2019.

\bibitem{jumper2021highly}
J.~Jumper, R.~Evans, A.~Pritzel, T.~Green, M.~Figurnov, O.~Ronneberger,
  K.~Tunyasuvunakool, R.~Bates, A.~{\v{Z}}{\'\i}dek, A.~Potapenko
  \emph{et~al.}, ``Highly accurate protein structure prediction with
  alphafold,'' \emph{Nature}, vol. 596, no. 7873, pp. 583--589, 2021.

\bibitem{physics}
\BIBentryALTinterwordspacing
A.~Sanchez{-}Gonzalez, J.~Godwin, T.~Pfaff, R.~Ying, J.~Leskovec, and P.~W.
  Battaglia, ``Learning to simulate complex physics with graph networks,''
  \emph{CoRR}, vol. abs/2002.09405, 2020. [Online]. Available:
  \url{https://arxiv.org/abs/2002.09405}
\BIBentrySTDinterwordspacing

\bibitem{978-3-642-02124-4_5}
B.~Mokbel, A.~Hasenfuss, and B.~Hammer, ``Graph-based representation of
  symbolic musical data,'' in \emph{Graph-Based Representations in Pattern
  Recognition}, A.~Torsello, F.~Escolano, and L.~Brun, Eds.\hskip 1em plus
  0.5em minus 0.4em\relax Berlin, Heidelberg: Springer Berlin Heidelberg, 2009,
  pp. 42--51.

\bibitem{SimonettaCOR18}
\BIBentryALTinterwordspacing
F.~Simonetta, F.~Carnovalini, N.~Orio, and A.~Rod{\`{a}}, ``Symbolic music
  similarity through a graph-based representation,'' in \emph{Proceedings of
  the Audio Mostly 2018 on Sound in Immersion and Emotion, Wrexham, United
  Kingdom, September 12-14, 2018}, S.~Cunningham and R.~Picking, Eds.\hskip 1em
  plus 0.5em minus 0.4em\relax {ACM}, 2018, pp. 26:1--26:7. [Online].
  Available: \url{https://doi.org/10.1145/3243274.3243301}
\BIBentrySTDinterwordspacing

\bibitem{jeong2019graph}
D.~Jeong, T.~Kwon, Y.~Kim, and J.~Nam, ``Graph neural network for music score
  data and modeling expressive piano performance,'' in \emph{International
  Conference on Machine Learning}.\hskip 1em plus 0.5em minus 0.4em\relax PMLR,
  2019, pp. 3060--3070.

\bibitem{abs-2208-14819}
\BIBentryALTinterwordspacing
E.~Karystinaios and G.~Widmer, ``Cadence detection in symbolic classical music
  using graph neural networks,'' \emph{CoRR}, vol. abs/2208.14819, 2022.
  [Online]. Available: \url{https://doi.org/10.48550/arXiv.2208.14819}
\BIBentrySTDinterwordspacing

\bibitem{10.2307/745814}
R.~D. Morris, ``{New Directions in the Theory and Analysis of Musical
  Contour},'' \emph{Music Theory Spectrum}, vol.~15, no.~2, pp. 205--228, 10
  1993.

\bibitem{szeto2006graph}
W.~M. Szeto and M.~H. Wong, ``A graph-theoretical approach for pattern matching
  in post-tonal music analysis,'' \emph{Journal of New Music Research},
  vol.~35, no.~4, pp. 307--321, 2006.

\bibitem{TruongOV20}
\BIBentryALTinterwordspacing
C.~Truong, L.~Oudre, and N.~Vayatis, ``Selective review of offline change point
  detection methods,'' \emph{Signal Process.}, vol. 167, 2020. [Online].
  Available: \url{https://doi.org/10.1016/j.sigpro.2019.107299}
\BIBentrySTDinterwordspacing

\bibitem{pelt}
R.~Killick, P.~Fearnhead, and I.~A. Eckley, ``Optimal detection of changepoints
  with a linear computational cost,'' \emph{Journal of the American Statistical
  Association}, vol. 107, no. 500, pp. 1590--1598, 2012.

\bibitem{musicaiz}
\BIBentryALTinterwordspacing
C.~Hernandez-Olivan and J.~R. Beltran, ``musicaiz: A python library for
  symbolic music generation, analysis and visualization,'' 2022. [Online].
  Available: \url{https://arxiv.org/abs/2209.07974}
\BIBentrySTDinterwordspacing

\bibitem{networkx}
A.~Hagberg, P.~Swart, and D.~S~Chult, ``Exploring network structure, dynamics,
  and function using networkx,'' Los Alamos National Lab.(LANL), Los Alamos, NM
  (United States), Tech. Rep., 2008.

\bibitem{raffel2014mir_eval}
C.~Raffel, B.~McFee, E.~J. Humphrey, J.~Salamon, O.~Nieto, D.~Liang, D.~P.
  Ellis, and C.~C. Raffel, ``mir\_eval: A transparent implementation of common
  mir metrics,'' in \emph{In Proceedings of the 15th International Society for
  Music Information Retrieval Conference, ISMIR}, 2014.

\bibitem{grill2015music}
T.~Grill and J.~Schl{\"{u}}ter, ``Music boundary detection using neural
  networks on spectrograms and self-similarity lag matrices,'' in \emph{23rd
  European Signal Processing Conference, {EUSIPCO} 2015, Nice, France, August
  31 - September 4, 2015}.\hskip 1em plus 0.5em minus 0.4em\relax {IEEE}, 2015,
  pp. 1296--1300.

\bibitem{bps}
\BIBentryALTinterwordspacing
T.~Chen and L.~Su, ``Functional harmony recognition of symbolic music data with
  multi-task recurrent neural networks,'' in \emph{Proceedings of the 19th
  International Society for Music Information Retrieval Conference, {ISMIR}
  2018, Paris, France, September 23-27, 2018}, E.~G{\'{o}}mez, X.~Hu,
  E.~Humphrey, and E.~Benetos, Eds., 2018, pp. 90--97. [Online]. Available:
  \url{http://ismir2018.ircam.fr/doc/pdfs/178\_Paper.pdf}
\BIBentrySTDinterwordspacing

\bibitem{GothamI19}
M.~Gotham and M.~Ireland, ``Taking form: {A} representation standard,
  conversion code, and example corpora for recording, visualizing, and studying
  analyses of musical form,'' in \emph{Proceedings of the 20th International
  Society for Music Information Retrieval Conference, {ISMIR} 2019, Delft, The
  Netherlands, November 4-8, 2019}, A.~Flexer, G.~Peeters, J.~Urbano, and
  A.~Volk, Eds., 2019, pp. 693--699.

\bibitem{ens2020mmm}
J.~Ens and P.~Pasquier, ``Mmm: Exploring conditional multi-track music
  generation with the transformer,'' \emph{arXiv preprint arXiv:2008.06048},
  2020.

\bibitem{augmentednet}
N.~N\'apoles~L\'opez, M.~Gotham, and I.~Fujinaga, ``{AugmentedNet: A Roman
  Numeral Analysis Network with Synthetic Training Examples and Additional
  Tonal Tasks},'' in \emph{{Proceedings of the 22nd International Society for
  Music Information Retrieval Conference}}.\hskip 1em plus 0.5em minus
  0.4em\relax ISMIR, Oct. 2021.

\end{thebibliography}

\end{document}